\documentclass[aps,pra,twocolumn,groupedaddress]{revtex4-2}

\usepackage{bm}

\usepackage{color,pstcol}

\usepackage{graphicx}

\newcommand{\Yqubitonesevspace}
{{\cal E}}
\newcommand{\yqubitoneindexstd}
{$\Yqubitoneindexstd$}
\newcommand{\Yqubitoneindexstd}
{j}

\newcommand{\Ytwoqubitwritereadtimeinterval}
{\tau}

\newcommand{\Yqubitbothtimeinitstatecoefindexequalstd}
{k}

\newcommand{\Yqubitbothtimefinalstatecoefindexindexstd}
{k}
\newcommand{\yopmix}
{$\Yopmix$}
\newcommand{\Yopmix}
{M}
\newcommand{\Yopmixbases}
{Q}
\newcommand{\Yopmixbasesimagin}
{\Yopmixbases_{\Ysqrtminusone}}

\newcommand{\Ysepsystdedicstateoutcoefnot}
{c}
\newcommand{\Ysepsystdedicstateoutcoefpart}
{\Ysepsystdedicstateoutcoefnot_{5}}

\newcommand{\Ysepsystdedicstateoutcoefprobnot}
{P}

\newcommand{\Ysepsystdedicstateoutcoefprobnotox}
{P}

\newcommand{\Ysepsystdedicopmixbasesonestateoutcoefprobzznot}
{\Ysepsystdedicstateoutcoefprobnot}

\newcommand{\Ysepsystdedicopmixbasesimaginonestateoutcoefprobzznot}
{\Ysepsystdedicstateoutcoefprobnot}
\newcommand{\ysepsystdedicopmixbasesimaginonestateoutcoefprobzzplusplus}
{$\Ysepsystdedicopmixbasesimaginonestateoutcoefprobzzplusplus$}
\newcommand{\Ysepsystdedicopmixbasesimaginonestateoutcoefprobzzplusplus}
{\Ysepsystdedicopmixbasesimaginonestateoutcoefprobzznot
_{1z}
( \Yopmixbasesimagin )
}

\newcommand{\Yindexstdforsepmixdiagel}
{k}
\newcommand{\yqubitnbarb}
{$\Yqubitnbarb$}
\newcommand{\Yqubitnbarb}
{Q}
\newcommand{\Yqubitindexstd}
{j}

\newcommand{\Yqubitnbarbstatecoefindexequalstd}
{k}
\newcommand{\ytwoqubitseqnb}
{$\Ytwoqubitseqnb$}
\newcommand{\Ytwoqubitseqnb}
{N}

\newcommand{\Ytwoqubitresultphaseinit}
{\Delta _I}

\newcommand{\Ytwoqubitresultphaseevolsin}
{v}

\newcommand{\Ytwoqubitresultphaseevolsinestim}
{\overline{\Ytwoqubitresultphaseevolsin}}

\newcommand{\Ytwoqubitresultphaseevolsinestimtwo}
{\widehat{\Ytwoqubitresultphaseevolsin}}

\newcommand{\Yparamqubitbothstateminusmodulusnot}
{q}
\newcommand{\Yparamqubitonestateminusmodulus}
{{\Yparamqubitbothstateminusmodulusnot}_1}
\newcommand{\Yparamqubittwostateminusmodulus}
{{\Yparamqubitbothstateminusmodulusnot}_2}
\newcommand{\Yparamqubitbothstateplusphasenot}
{\theta}
\newcommand{\Yparamqubitonestateplusphase}
{\Yparamqubitbothstateplusphasenot_1}
\newcommand{\Yparamqubittwostateplusphase}
{\Yparamqubitbothstateplusphasenot_2}
\newcommand{\Yparamqubitbothstateminusphasenot}
{\phi}
\newcommand{\Yparamqubitonestateminusphase}
{\Yparamqubitbothstateminusphasenot_1}
\newcommand{\Yparamqubittwostateminusphase}
{\Yparamqubitbothstateminusphasenot_2}
\newcommand{\Ymagfieldnot}
{B}
\newcommand{\Ymagfieldvec}
{\overrightarrow{\Ymagfieldnot}}
\newcommand{\Yhamiltonfieldscale}
{G}
\newcommand{\yopdensity}
{$\Yopdensity$}
\newcommand{\Yopdensity}
{\rho}

\newcommand{\yphysicalquantitystdone}
{$\Yphysicalquantitystdone$}
\newcommand{\Yphysicalquantitystdone}
{A}

\newcommand{\yobservableopstdone}
{$\Yobservableopstdone$}
\newcommand{\Yobservableopstdone}
{\hat{\Yphysicalquantitystdone}}

\newcommand{\Yobservableopstdoneindexelone}
{
k
}

\newcommand{\Yobservableopstdoneindexeltwo}
{
\ell
}

\newcommand{\Yobservableopstdoneelnot}
{a}

\newcommand{\yobservableopstdoneelindexstdoneindexstdtwo}
{$\Yobservableopstdoneelindexstdoneindexstdtwo$}
\newcommand{\Yobservableopstdoneelindexstdoneindexstdtwo}
{
\Yobservableopstdoneelnot
_{
\Yobservableopstdoneindexelone
\Yobservableopstdoneindexeltwo
}
}

\newcommand{\Ysqrtminusone}
{i}

\newcommand{\Ytrace}
{\mathrm{Tr}}

\newcommand{\ywritereadonestatenb}
{$\Ywritereadonestatenb$}
\newcommand{\Ywritereadonestatenb}
{K}
\newcommand{\yeqdefstatepuredeterm}
{
\begin{equation}
\Yketdeterm
=
\sum
_{
\Yindexstdforketbasis
=
0
}
^
{
\Yketspacedim
-
1
}
\Yketbasiscoefdetermindexstd
\Yketbasisindexstd
\label{eq-statepuredeterm}
\end{equation}
}
\newcommand{\yeqdefstatepurerand}
{
\begin{equation}
\Yketrand
=
\sum
_{
\Yindexstdforketbasis
=
0
}
^
{
\Yketspacedim
-
1
}
\Yketbasiscoefrandindexstd
\Yketbasisindexstd
\label{eq-statepurerand}
\end{equation}
}
\newcommand{\yeqdefstatepurerandval}
{
\begin{equation}
\Yketrandval
=
\sum
_{
\Yindexstdforketbasis
=
0
}
^
{
\Yketspacedim
-
1
}
\Yketbasiscoefrandindexstdval
\Yketbasisindexstd
.
\label{eq-statepurerandval}
\end{equation}
}

\newcommand{\yeqstatepurerandvalnormalized}
{
\begin{equation}
\sum
_{
\Yindexstdforketbasis
=
0
}
^
{
\Yketspacedim
-
1
}
|
\Yketbasiscoefrandindexstdval
|
^2
=
1
\label{eq-statepurerandval-normalized}
\end{equation}
}
\newcommand{\yeqdefstatepurerandonequbit}
{
\begin{equation}
\Yketrand
=
\Yketbasiscoefrandindexzeroamongtwomod
\Yketbasisindexzero
+
\sqrt{
1
-
\Yketbasiscoefrandindexzeroamongtwomod
^2
}
e^{
\Ysqrtminusone
\Yketbasiscoefrandindexoneamongtwophase
}
\Yketbasisindexone
\label{eq-statepurerandonequbit}
\end{equation}
}
\newcommand{\yeqstatepurerandvalmeasprob}
{
\begin{equation}
\Yketbasisprobrandindexstdval
=
|
\Yketbasiscoefrandindexstdval
|
^2
.
\label{eq-statepurerandval-measprob}
\end{equation}
}
\newcommand{\yeqstatepurerandmeasprob}
{
\begin{equation}
\Yketbasisprobrandindexstd
=
|
\Yketbasiscoefrandindexstd
|
^2
.
\label{eq-statepurerand-measprob}
\end{equation}
}
\newcommand{\yeqstatepurerandmeasprobonequbit}
{
\begin{eqnarray}
\Yketbasisprobrandindexzero
&
=
&
\Yketbasiscoefrandindexzeroamongtwomod
^2
\label{eq-statepurerand-measprob-onequbit-eq-one}
\\
\Yketbasisprobrandindexone
&
=
&
1
-
\Yketbasiscoefrandindexzeroamongtwomod
^2
=
1
-
\Yketbasisprobrandindexzero
.
\label{eq-statepurerand-measprob-onequbit}
\end{eqnarray}
}
\newcommand{\yeqdefdensoprandstateelindexstdindexstdtwo}
{
\begin{equation}
\Ydensoprandstateelindexstdindexstdtwo
=
E\{\Yketbasiscoefrandindexstd
\Yketbasiscoefrandindexstdtwo
^*\}
.
\label{eq-defdensoprandstateelindexstdindexstdtwo}
\end{equation}
}
\newcommand{\yeqdefdensoprandstateelindexstdindexstd}
{
\begin{equation}
\Ydensoprandstateelindexstdindexstd
=
E
\{
|
\Yketbasiscoefrandindexstd
|
^2
\}
=
E
\{
\Yketbasisprobrandindexstd
\}
.
\label{eq-defdensoprandstateelindexstdindexstd}
\end{equation}
}
\newcommand{\yeqobservablemeanpurestate}
{
\begin{equation}
E
\{
\Yphysicalquantitystdone
\}
_{
\Yketdeterm
}
=
\Ybradeterm
\Yobservableopstdone
\Yketdeterm
\label{eq-observablemeanpurestate}
\end{equation}
}
\newcommand{\yeqobservablemeanmixedstate}
{
\begin{equation}
E
\{
\Yphysicalquantitystdone
\}
_{
\Yopdensity
}
=
\Ytrace(\Yopdensity\Yobservableopstdone)
\label{eq-observablemeanmixedstate}
\end{equation}
}
\newcommand{\yeqdetermcoefpurestatedensityop}
{
\begin{equation}
\Ydensopmixedstate
= 
\Yketdeterm
\Ybradeterm
\label{eq-determcoefpurestatedensityop}
\end{equation}
}
\newcommand{\yeqdetermcoefpurestatedensitymatrixel}
{
\begin{equation}
\Ydensopmixedstateelindexstdindexstdtwo
=
\Yketbasiscoefdetermindexstd
\Yketbasiscoefdetermindexstdtwo
^*
\label{eq-determcoefpurestatedensitymatrixel}
\end{equation}
}
\newcommand{\yeqrandstatedensitymatrixelapproachtwoout}
{
\begin{equation}
\Ydensoprandstateelindexstdindexstdtwoapproachtwoout
=
\Yketbasiscoefrandindexstdval
\Yketbasiscoefrandindexstdtwoval
^*
\label{eq-randstatedensitymatrixelapproachtwoout}
\end{equation}
}
\newcommand{\yeqrandstatedensitymatrixelapproachtworv}
{
\begin{equation}
\Ydensoprandstateelindexstdindexstdtwoapproachtwo
=
\Yketbasiscoefrandindexstd
\Yketbasiscoefrandindexstdtwo
^*
.
\label{eq-randstatedensitymatrixelapproachtwoorv}
\end{equation}
}
\newcommand{\yeqphysicalquantitystdonemeanforketrandval}
{
\begin{equation}
E
\{
\Yphysicalquantitystdone
\}
_{\Yketrandval}
=
\sum_{\Yobservableopstdoneindexelone}
\sum_{\Yobservableopstdoneindexeltwo}
\Yketbasiscoefdetermnot
_{
\Yobservableopstdoneindexelone
}
(
\Yqipout
)
^*
\Yketbasiscoefdetermnot
_{
\Yobservableopstdoneindexeltwo
}
(
\Yqipout
)
\Yobservableopstdoneelindexstdoneindexstdtwo
.
\label{eq-physicalquantitystdonemeanforketrandval}
\end{equation}
}
\newcommand{\yeqphysicalquantitystdonemeanforketrand}
{
\begin{equation}
E
\{
\Yphysicalquantitystdone
\}
_{\Yketrand}
=
\sum_{\Yobservableopstdoneindexelone}
\sum_{\Yobservableopstdoneindexeltwo}
E
\{
\boldsymbol{\mathbf{
\Yketbasiscoefdetermnot
_{
\Yobservableopstdoneindexelone
}
}}
^*
\boldsymbol{\mathbf{
\Yketbasiscoefdetermnot
_{
\Yobservableopstdoneindexeltwo
}
}}
\}
\Yobservableopstdoneelindexstdoneindexstdtwo
.
\label{eq-physicalquantitystdonemeanforketrand}
\end{equation}
}

\newcommand{\yeqphysicalquantitystdonefuncstdonemeanforketrandversiontwo}
{
\begin{equation}
E
\{
\Yqipfuncstdone
(
\Yphysicalquantitystdone
)
\}
_{\Yketrand}
=
\sum_{\Yobservableopstdoneindexelone}
\sum_{\Yobservableopstdoneindexeltwo}
E
\{
\boldsymbol{\mathbf{
\Yketbasiscoefdetermnot
_{
\Yobservableopstdoneindexelone
}
}}
^*
\boldsymbol{\mathbf{
\Yketbasiscoefdetermnot
_{
\Yobservableopstdoneindexeltwo
}
}}
\}
g
_{
\Yobservableopstdoneindexelone
\Yobservableopstdoneindexeltwo
}
\label{eq-physicalquantitystdonefuncstdonemeanforketrandversiontwo}
\end{equation}
}
\newcommand{\yeqketbasisprobrandindexstdjointmoment}
{
\begin{equation}
E
\{
\prod
_{
\Yindexstdforketbasis
\in
I
}
(
\Yketbasisprobrandindexstd
)
^{
m
_{
\Yindexstdforketbasis
}
}
\}
\label{eq-ketbasisprobrandindexstdjointmoment}
\end{equation}
}

\newcommand{\yqipspace}{$\Yqipspace$}
\newcommand{\Yqipspace}{\Omega}

\newcommand{\yqipout}{$\Yqipout$}
\newcommand{\Yqipout}{\alpha}

\newcommand{\yqipoutnb}{$\Yqipoutnb$}
\newcommand{\Yqipoutnb}
{L}

\newcommand{\yqipoutprob}{$\Yqipoutprob$}
\newcommand{\Yqipoutprob}{P(\Yqipout)}

\newcommand{\yqipRV}{$\YqipRV$}
\newcommand{\YqipRV}
{\mathbf{X}}

\newcommand{\yqipRVtwo}{$\YqipRVtwo$}
\newcommand{\YqipRVtwo}{\mathbf{Y}}

\newcommand{\yqipRVval}{$\YqipRVval$}
\newcommand{\YqipRVval}{X ( \Yqipout )}

\newcommand{\yketspacedim}{$\Yketspacedim$}
\newcommand{\Yketspacedim}{d}

\newcommand{\yindexstdforketbasis}{$\Yindexstdforketbasis$}
\newcommand{\Yindexstdforketbasis}{k}

\newcommand{\yindexstdtwoforketbasis}{$\Yindexstdtwoforketbasis$}
\newcommand{\Yindexstdtwoforketbasis}{\ell}

\newcommand{\Yketbasisindexzero}
{
|
0
\rangle
}

\newcommand{\Yketbasisindexone}
{
|
1
\rangle
}

\newcommand{\yketbasisindexstd}{$\Yketbasisindexstd$}
\newcommand{\Yketbasisindexstd}
{
|
\Yindexstdforketbasis
\rangle
}

\newcommand{\Yketbasiscoefdetermnot}
{c}

\newcommand{\Yketbasiscoefdetermindexzero}
{
\Yketbasiscoefdetermnot
_{
0
}
}

\newcommand{\Yketbasiscoefdetermindexzeroamongtwomod}
{
r
}

\newcommand{\Yketbasiscoefdetermindexoneamongtwophase}
{
\phi
}

\newcommand{\yketbasiscoefdetermindexstd}{$\Yketbasiscoefdetermindexstd$}
\newcommand{\Yketbasiscoefdetermindexstd}
{
\Yketbasiscoefdetermnot
_{
\Yindexstdforketbasis
}
}

\newcommand{\Yketbasiscoefdetermindexstdtwo}
{
\Yketbasiscoefdetermnot
_{
\Yindexstdtwoforketbasis
}
}

\newcommand{\Yketbasiscoefdetermindexlast}
{
\Yketbasiscoefdetermnot
_{
\Yketspacedim -1
}
}

\newcommand{\Yketdetermnot}{\psi}

\newcommand{\yketdeterm}{$\Yketdeterm$}
\newcommand{\Yketdeterm}{| \Yketdetermnot \rangle}

\newcommand{\Ybradeterm}{\langle \Yketdetermnot |}

\newcommand{\Yketbasisprobdetermnot}
{p}

\newcommand{\Yketbasisprobdetermindexzero}
{
\Yketbasisprobdetermnot
_{
0
}
}

\newcommand{\Yketbasisprobdetermindexone}
{
\Yketbasisprobdetermnot
_{
1
}
}

\newcommand{\Yketbasisprobdetermindexstd}
{
\Yketbasisprobdetermnot
_{
\Yindexstdforketbasis
}
}

\newcommand{\Ydensopmixedstate}{\rho}

\newcommand{\Ydensopmixedstateelindexstdindexstdtwo}
{
\Ydensopmixedstate
_{
\Yindexstdforketbasis
\Yindexstdtwoforketbasis
}
}

\newcommand{\yketbasiscoefrandindexzero}{$\Yketbasiscoefrandindexzero$}
\newcommand{\Yketbasiscoefrandindexzero}
{
\boldsymbol{\mathbf{
\Yketbasiscoefdetermnot
_{
0
}
}}
}

\newcommand{\yketbasiscoefrandindexzeroval}{$\Yketbasiscoefrandindexzeroval$}
\newcommand{\Yketbasiscoefrandindexzeroval}
{
\Yketbasiscoefdetermindexzero
(
\Yqipout
)
}

\newcommand{\yketbasiscoefrandindexzeroamongtwomod}{$\Yketbasiscoefrandindexzeroamongtwomod$}
\newcommand{\Yketbasiscoefrandindexzeroamongtwomod}
{
\boldsymbol{\mathbf{\Yketbasiscoefdetermindexzeroamongtwomod}}
}

\newcommand{\yketbasiscoefrandindexzeroamongtwovalmod}{$\Yketbasiscoefrandindexzeroamongtwovalmod$}
\newcommand{\Yketbasiscoefrandindexzeroamongtwovalmod}
{
\Yketbasiscoefdetermindexzeroamongtwomod
(
\Yqipout
)
}

\newcommand{\yketbasiscoefrandindexzeroamongtwomoddistribboundlow}{$\Yketbasiscoefrandindexzeroamongtwomoddistribboundlow$}
\newcommand{\Yketbasiscoefrandindexzeroamongtwomoddistribboundlow}
{
\Yketbasiscoefdetermindexzeroamongtwomod
_{1}
}

\newcommand{\yketbasiscoefrandindexzeroamongtwomoddistribboundhigh}{$\Yketbasiscoefrandindexzeroamongtwomoddistribboundhigh$}
\newcommand{\Yketbasiscoefrandindexzeroamongtwomoddistribboundhigh}
{
\Yketbasiscoefdetermindexzeroamongtwomod
_{2}
}

\newcommand{\yketbasiscoefrandindexone}{$\Yketbasiscoefrandindexone$}
\newcommand{\Yketbasiscoefrandindexone}
{
\boldsymbol{\mathbf{
\Yketbasiscoefdetermnot
_{
1
}
}}
}

\newcommand{\yketbasiscoefrandindexoneamongtwophase}{$\Yketbasiscoefrandindexoneamongtwophase$}
\newcommand{\Yketbasiscoefrandindexoneamongtwophase}
{
\boldsymbol{\Yketbasiscoefdetermindexoneamongtwophase}
}

\newcommand{\yketbasiscoefrandindexoneamongtwophasedistribbound}{$\Yketbasiscoefrandindexoneamongtwophasedistribbound$}
\newcommand{\Yketbasiscoefrandindexoneamongtwophasedistribbound}
{
B
_{
\Yketbasiscoefrandindexoneamongtwophase
}
}

\newcommand{\yketbasiscoefrandindexstd}{$\Yketbasiscoefrandindexstd$}
\newcommand{\Yketbasiscoefrandindexstd}
{
\boldsymbol{\mathbf{
\Yketbasiscoefdetermnot
_{
\Yindexstdforketbasis
}
}}
}

\newcommand{\yketbasiscoefrandindexstdval}{$\Yketbasiscoefrandindexstdval$}
\newcommand{\Yketbasiscoefrandindexstdval}
{
\Yketbasiscoefdetermindexstd
(
\Yqipout
)
}

\newcommand{\yketbasiscoefrandindexstdmod}{$\Yketbasiscoefrandindexstdmod$}
\newcommand{\Yketbasiscoefrandindexstdmod}
{
\boldsymbol{\mathbf{
r
_{
\Yindexstdforketbasis
}
}}
}

\newcommand{\yketbasiscoefrandindexstdphase}{$\Yketbasiscoefrandindexstdphase$}
\newcommand{\Yketbasiscoefrandindexstdphase}
{
\boldsymbol{\mathbf{
\phi
_{
\Yindexstdforketbasis
}
}}
}

\newcommand{\Yketbasiscoefrandindexstdtwo}
{
\boldsymbol{\mathbf{
\Yketbasiscoefdetermnot
_{
\Yindexstdtwoforketbasis
}
}}
}

\newcommand{\Yketbasiscoefrandindexstdtwoval}
{
\Yketbasiscoefdetermindexstdtwo
(
\Yqipout
)
}

\newcommand{\Yketbasiscoefrandindexlastval}
{
\Yketbasiscoefdetermindexlast
(
\Yqipout
)
}

\newcommand{\Yketrandnot}{\boldsymbol{\Yketdetermnot}}

\newcommand{\yketrand}{$\Yketrand$}
\newcommand{\Yketrand}{| \Yketrandnot \rangle}

\newcommand{\yketrandval}{$\Yketrandval$}
\newcommand{\Yketrandval}{| 
\Yketdetermnot
(
\Yqipout
)
\rangle
}

\newcommand{\yketrandprocessin}{$\Yketrandprocessin$}
\newcommand{\Yketrandprocessin}{| 
\boldsymbol{\Yketdetermnot _{in}}
\rangle}

\newcommand{\yketrandprocessout}{$\Yketrandprocessout$}
\newcommand{\Yketrandprocessout}{| 
\boldsymbol{\Yketdetermnot _{out}}
\rangle}

\newcommand{\yketbasisprobrandindexzero}{$\Yketbasisprobrandindexzero$}
\newcommand{\Yketbasisprobrandindexzero}
{
\boldsymbol{\mathbf{
\Yketbasisprobdetermindexzero
}}
}

\newcommand{\Yketbasisprobrandindexone}
{
\boldsymbol{\mathbf{
\Yketbasisprobdetermindexone
}}
}

\newcommand{\yketbasisprobrandindexstd}{$\Yketbasisprobrandindexstd$}
\newcommand{\Yketbasisprobrandindexstd}
{
\boldsymbol{\mathbf{
\Yketbasisprobdetermindexstd
}}
}

\newcommand{\yketbasisprobrandindexstdval}{$\Yketbasisprobrandindexstdval$}
\newcommand{\Yketbasisprobrandindexstdval}
{
\Yketbasisprobdetermindexstd
(
\Yqipout
)
}

\newcommand{\Ydensoprandstate}{\rho}

\newcommand{\ydensoprandstateelindexstdindexstd}{$\Ydensoprandstateelindexstdindexstd$}
\newcommand{\Ydensoprandstateelindexstdindexstd}
{
\Ydensoprandstate
_{
\Yindexstdforketbasis
\Yindexstdforketbasis
}
}

\newcommand{\Ydensoprandstateelindexstdindexstdtwo}
{
\Ydensoprandstate
_{
\Yindexstdforketbasis
\Yindexstdtwoforketbasis
}
}

\newcommand{\Ydensoprandstateapproachtwonot}{\tilde{\rho}}

\newcommand{\ydensoprandstateapproachtwoout}{$\Ydensoprandstateapproachtwoout$}
\newcommand{\Ydensoprandstateapproachtwoout}
{
\Ydensoprandstateapproachtwonot
(
\Yqipout
)
}

\newcommand{\Ydensoprandstateelindexstdindexstdtwoapproachtwoout}
{
\Ydensoprandstateapproachtwonot
_{
\Yindexstdforketbasis
\Yindexstdtwoforketbasis
}
(
\Yqipout
)
}

\newcommand{\ydensoprandstateapproachtwo}{$\Ydensoprandstateapproachtwo$}
\newcommand{\Ydensoprandstateapproachtwo}
{\boldsymbol{\Ydensoprandstateapproachtwonot}}

\newcommand{\Ydensoprandstateelindexstdindexstdtwoapproachtwo}
{
\Ydensoprandstateapproachtwo
_{
\Yindexstdforketbasis
\Yindexstdtwoforketbasis
}
}

\newcommand{\yqipfuncstdone}{$\Yqipfuncstdone$}
\newcommand{\Yqipfuncstdone}{g}

\newcommand{\Yopmixparamnot}
{v}
\newcommand{\yopmixparamone}
{$\Yopmixparamone$}
\newcommand{\Yopmixparamone}
{\Yopmixparamnot_{1}}
\newcommand{\yopmixparamtwo}
{$\Yopmixparamtwo$}
\newcommand{\Yopmixparamtwo}
{\Yopmixparamnot_{2}}
\newcommand{\yopmixparamthree}
{$\Yopmixparamthree$}
\newcommand{\Yopmixparamthree}
{\Yopmixparamnot_{3}}
\newcommand{\yopmixparamfour}
{$\Yopmixparamfour$}
\newcommand{\Yopmixparamfour}
{\Yopmixparamnot_{4}}

\newcommand{\Yketbasisprobrandindexzeropowercoeftwo}
{a_2}

\newcommand{\Yketbasisprobrandindexzeropowercoefone}
{a_1}

\newcommand{\Yketbasisprobrandindexzeropowercoefzero}
{a_0}

\newcommand{\Yketbasisprobrandindexzeropowercoefsubtermone}
{b_1}

\newcommand{\Yketbasisprobrandindexzeropowercoefsubtermtwo}
{b_2}

\newcommand{\yopmixparamthreesolindetermsignone}
{$\Yopmixparamthreesolindetermsignone$}
\newcommand{\Yopmixparamthreesolindetermsignone}
{\epsilon_1}
\newcommand{\yopmixparamthreesolindetermsigntwo}
{$\Yopmixparamthreesolindetermsigntwo$}
\newcommand{\Yopmixparamthreesolindetermsigntwo}
{\epsilon_2}
\newcommand{\yopmixparamfoursolindetermsign}
{$\Yopmixparamfoursolindetermsign$}
\newcommand{\Yopmixparamfoursolindetermsign}
{\epsilon_3}

\newcommand{\ytextmodifartitionehundredsixtyfivevonestepone}[1]
{#1}

\begin{document}


\title{%
Beyond the density operator
and
$
\boldsymbol{Tr( \rho \hat{A} )}
$%
:
Exploiting 
the
higher-order statistics
of random-coefficient pure 
states
for 
quantum
information processing%
}


\author{%
{Yannick Deville}%
}
\email[]{yannick.deville@irap.omp.eu}
\affiliation{%
{%
{%
Universit\'e de Toulouse}%
,
UPS, CNRS, CNES,
OMP,\\
IRAP
(Institut de Recherche en Astrophysique et Plan\'etologie),
F-31400
Toulouse,
France%
}
}

\author{%
{Alain Deville}}
\email[]{alain.deville@univ-amu.fr}
\affiliation{%
{%
Aix-Marseille Universit\'e,
CNRS%
,
IM2NP UMR 7334%
,
F-%
13397
Marseille, France%
}
}


\date{\today}

\begin{abstract}
Two types of states are 
widely used
in quantum mechanics,
namely (deterministic-coefficient) pure states and statistical mixtures.
A density operator
can be associated with each
of them.
In this paper, we address
a
third type of states,
that we 
previously
introduced 
in a more restricted framework.
These states generalize pure ones
by replacing 
each of 
their deterministic 
ket
coefficients by
a random variable.
We therefore call them
Random-Coefficient Pure States, or RCPS.
We 
here
analyze 
their properties
and their relationships with
both types of usual states.
We show that RCPS contain much richer information than the density
operator 
and mean of observables
that we 
associate
with them.
This occurs because
the latter operator 
only exploits 
the second-order statistics
of the random state coefficients,
whereas their higher-order statistics contain additional information.
That information can
be accessed in practice
with the 
multiple-preparation 
procedure
that we propose for RCPS,
by
using
second-order and higher-order statistics of
associated random probabilities of measurement outcomes
(we also discuss
our single-preparation 
procedure).
Exploiting these higher-order statistics
opens the way to a very general approach for performing
advanced quantum information processing tasks.
We illustrate the relevance of this approach
with a generic
example, 
dealing with the 
estimation
of parameters of a quantum process
and thus related to quantum process tomography.
This parameter estimation is
performed in the non-blind (i.e. supervised) or blind
(i.e. unsupervised) mode.
For the considered type of measurements,
we show that this problem cannot be solved by using only the
density operator 
$
\rho
$
of
an RCPS
and the associated mean value
$
Tr( \rho \hat{A} )
$
of 
the operator
$
\hat{A}
$
that
corresponds to
the considered
physical quantity.
In contrast,
we succeed in solving this problem
by 
exploiting a fourth-order
statistical parameter of state coefficients, in addition to
second-order statistics.
Numerical tests validate this 
result
and show that the proposed method
yields accurate parameter estimation for the considered
number of state preparations.
\end{abstract}


\maketitle

\section{Introduction}
\label{sec-intro}
Two types of states are 
widely
used
in quantum mechanics,
namely pure states 
(with deterministic coefficients: see below) and mixed states, i.e.
statistical mixtures, the latter being a superset of the
former.
Due to our needs for new classes of quantum information
processing (QIP) methods, 
in
\cite{amoi5-31}
we introduced 
a third
approach, 
based on the concept 
that we 
then called ``random pure states'',
and that is hereafter more precisely referred to as
Random-Coefficient 
Pure States
and
abbreviated as
RCPS.

We previously used these RCPS to perform various QIP tasks based on
blind adaptation/estimation, i.e. unsupervised
quantum machine learning
\cite{amoi-ieee-tqe-2021}.
These tasks are
Blind Quantum Source Separation
(BQSS, 
introduced in \cite{amoi5-31};
see also e.g.
\cite{amoi6-18,amoi6-42,amoi-ieee-tqe-2021}),
Blind Quantum Process Tomography
(BQPT, 
introduced in \cite{amoi6-46}; see also e.g. \cite{amoi6-118}),
Blind Hamiltonian Parameter Estimation
(BHPE, introduced in
\cite{amoi-ieee-tqe-2021})
and 
\ytextmodifartitionehundredsixtyfivevonestepone{other QIP tasks}
\cite{amoi-ieee-tqe-2021}.
Beyond the above practical QIP methods, we started to
investigate more fundamental aspects of RCPS in \cite{amoi6-67}:
we showed how these states can be physically
implemented and we briefly commented about their
relationship with the concept of density operator.
We addressed the latter topic 
in a much more detailed way
very recently in 
\cite{amoi-arxiv-2022-adeville-artiti163v2}.
This 
especially
showed that, starting from an RCPS, 
one can associate a density operator with it.

In this paper, we proceed much further in the investigation
of RCPS.
In Section
\ref{sec-rcps-def-et-autres},
we first provide a general definition of these states,
beyond their specific versions 
considered in our
above-mentioned
application-driven papers.
We 
then 
analyze various features 
of these states and show their potential
for QIP,
as compared with 
more standard
approaches.
We 
especially
explain how one may try to handle RCPS
by adapting
the
standard 
practice in
quantum mechanics, which is
based on defining other states
(namely mixed
ones)
by a density operator
\yopdensity\
and using the 
mean value
$\Ytrace(\Yopdensity\Yobservableopstdone)$
of a physical quantity
(i.e. observable)
\yphysicalquantitystdone\
represented by an operator
\yobservableopstdone.
We 
prove that this standard approach
does not allow one to access all the information that is
present in 
an RCPS.
That information can
indeed be accessed, by using measurements
and the associated statistics
of the moduli of the random ket coefficients
of that state.
\textit{A major result of this paper is thus that
certain QIP tasks
cannot be 
carried out
by only resorting to 
the standard approach to quantum mechanics,
whereas they can
be performed by exploiting the 
higher-order
statistics of 
the random coefficients of
an RCPS.}
In Section
\ref{sec-appli},
we illustrate this phenomenon
with a generic example,
dealing with 
non-blind or blind
quantum parameter estimation 
and 
related to 
(B)QPT
and
(B)HPE.
In Section
\ref{sec-compare-mixed-state},
we focus on the
discrete version of
RCPS
and
analyze their connections with
usual mixed states, 
as defined by
von Neumann.
Relationships with other works from the literature,
that are more or less
connected with RCPS
and their higher-order statistics, are
then discussed in Section
\ref{sec-other-works-random-states}.
Finally, we draw conclusions from this investigation in Section
\ref{sec-concl}.
\section{%
Definition and
features of
random-coefficient pure states (RCPS)}
\label{sec-rcps-def-et-autres}
\subsection{Definition of an RCPS}
\label{sec-rcps-def}
First considering the classical framework, 
the following concepts 
should be kept in mind.
Beyond a scalar deterministic (i.e. fixed) value
$X$,
a
random variable (RV) may
be defined
as a function
\yqipRV\
whose scalar
value
\yqipRVval\
depends on an outcome
\yqipout\
of the considered probability space
\yqipspace.
That outcome 
\yqipout\
is randomly drawn and, once selected, it
completely
defines the corresponding
(complex or real) value
\yqipRVval\
of
\yqipRV.
One may thus e.g. model an experiment where 
a die is cast,
each of its faces corresponds to an outcome
\yqipout,
and the user's numerical gain
\yqipRVval\
associated with each given face
\yqipout\
in a game
is fixed.
More generally, a random vector is a vector whose components are
RV,
i.e. all their values are fixed by the considered single outcome
\yqipout.

Now moving to the quantum framework,
the simplest states considered in the literature,
called pure
states, are deterministic:
such a state may be defined as a ket
\yeqdefstatepuredeterm
where the kets
\yketbasisindexstd\
form 
an orthonormal basis of the considered 
\yketspacedim-dimensional
space 
(with
$\Yketspacedim = 2^{\Yqubitnbarb}$
for
\yqubitnbarb\ qubits)
and
the corresponding complex-valued coefficients
\yketbasiscoefdetermindexstd\
are fixed 
for a given state
\yketdeterm.
In our above-mentioned papers, we extended that concept to
\textit{random-coefficient} pure states, or RCPS.
Such a state may be defined as a ket
\yeqdefstatepurerand
where
the complex-valued coefficients
\yketbasiscoefrandindexstd\
are 
RV, i.e. they depend on a randomly drawn outcome
\yqipout.
Once a single
\yqipout\
has been selected,
all corresponding coefficient values
\yketbasiscoefrandindexstdval\
are fixed,
as in a classical random vector.
A given outcome
\yqipout\
thus yields
a fixed, i.e. deterministic-coefficient, pure state
\yeqdefstatepurerandval

Such RCPS
\yketrand\
and
their
realizations
\yketrandval\
can actually be faced in practice.
For instance,
in \cite{amoi6-67},
we showed how to create an RCPS
for a single electron spin 1/2,
placed in a 
Stern-Gerlach
device
with a randomly drawn direction for the magnetic field.
A second example is introduced here for quantum communications.
In this scenario, the
receiver gets a pure state with coefficient values that he does not
know 
in advance, because he does not know which data were used by the emitter
to prepare the pure state that he sent.
The receiver may then describe the coefficients of the received
pure state with RV
\yketbasiscoefrandindexstd.

Whatever the considered RCPS,
the coefficients
\yketbasiscoefrandindexstdval\
of each
state realization
(\ref{eq-statepurerandval})
have the same
constraints as those of usual, i.e. deterministic-coefficient, pure states
(\ref{eq-statepuredeterm}):
the state 
\yketrandval\
is normalized, so that
\yeqstatepurerandvalnormalized
and \yketrandval\ is defined up to a global phase factor, so that
\yketbasiscoefrandindexzeroval\ may be restricted to a real non-negative
value
\yketbasiscoefrandindexzeroamongtwovalmod.
In particular,
setting
$\Yketspacedim = 2$ in the above equations,
an RCPS of a single qubit 
reads
\yeqdefstatepurerandonequbit
where
\yketbasiscoefrandindexzeroamongtwomod\
and
\yketbasiscoefrandindexoneamongtwophase\
are real-valued
RV
and
\yketbasiscoefrandindexzeroamongtwomod\
is non-negative.
\subsection{RCPS preparation and measurements}
\label{sec-prepar-measur}
Information about deterministic-coefficient or random-coefficient pure states can be
extracted by means of measurements.
For a given deterministic-coefficient
pure state
(\ref{eq-statepuredeterm})
or
(\ref{eq-statepurerandval}),
one may 
first 
use measurements
in the computational basis
$\{ \Yketbasisindexstd \}$,
which
e.g. consists of 
measuring 
the
$s_z$
spin component for an
electron spin 1/2
whose state is expressed in the standard basis.
The results of these measurements have a random nature,
but their possible values and 
the probabilities of these values
are fixed
for a given deterministic-coefficient pure state:
for state
(\ref{eq-statepurerandval}),
the probability of the result associated with the basis vector
\yketbasisindexstd\
is
\yeqstatepurerandvalmeasprob
Estimates of these probabilities may be obtained,
especially by
preparing
\ywritereadonestatenb\
copies of
\yketrandval,
performing one 
(possibly multiqubit)
measurement per copy
and computing the sample frequencies of all possible
measurement results over all these state copies
\cite{amq30official,amq98}.

Now consider a \textit{random-coefficient} pure state
\yketrand\ defined by
(\ref{eq-statepurerand}).
For any given basis vector
\yketbasisindexstd,
the probability
\yketbasisprobrandindexstdval\
depends on the randomly drawn outcome
\yqipout,
so that this type of probability itself becomes random-valued!
It defines an RV, that is denoted as
\yketbasisprobrandindexstd\
and that may be expressed as
\yeqstatepurerandmeasprob
For instance, for the single-qubit RCPS
(\ref{eq-statepurerandonequbit}),
this yields
\yeqstatepurerandmeasprobonequbit

Measurements may be used in a two-level procedure 
to extract information about 
an RCPS 
defined by (\ref{eq-statepurerand}).
At the higher level,
\ytwoqubitseqnb\
values of the set of coefficients
$
\{
\Yketbasiscoefrandindexzeroval
,
\dots
,
\Yketbasiscoefrandindexlastval
\}
$
associated with an outcome
\yqipout\
are
randomly drawn.
This
yields 
\ytwoqubitseqnb\
deterministic-coefficient states
\yketrandval\
defined by
(\ref{eq-statepurerandval}).
Then, at the lower level, for each such state
\yketrandval,
one uses
\ywritereadonestatenb\
copies of
\yketrandval\
to estimate
all
\yketbasisprobrandindexstdval\
as described above for deterministic-coefficient pure states.
For any index
\yindexstdforketbasis,
the overall set of 
\ytwoqubitseqnb\
estimates of
\yketbasisprobrandindexstdval\
thus obtained yields an estimate of the
statistical distribution
(i.e. law)
of the RV
\yketbasisprobrandindexstd.
One may then
e.g. derive the corresponding histogram,
which is an estimate of the
probability density function
(pdf) of
\yketbasisprobrandindexstd.

The above 
procedure involves randomness at \textit{two} levels,
instead of one level
for usual (i.e. deterministic-coefficient) pure states:
a) in the selection
of
the set of coefficients
$
\{
\Yketbasiscoefrandindexzeroval
,
\dots
,
\Yketbasiscoefrandindexlastval
\}
$,
i.e. 
in the selection
of an outcome
\yqipout,
and 
b) in the result 
provided by
a single (possibly multiqubit) measurement
performed for a given, i.e. deterministic-coefficient,  state.
In our previous papers, we first called that approach
the ``Repeated Write/Read'' or RWR approach,
with ``write'' referring to state preparation and
``read'' referring to measurements
(see e.g.
\cite{amoi5-31,amoi6-18}).
We then called it the ```multiple-preparation'' (per state
\yketrandval) approach
\cite{amoi-ieee-tqe-2021},
as opposed to the ``single-preparation
approach'' that we later proposed in
\cite{amoi6-104,amoi6-118,amoi-ieee-tqe-2021}
and that is considered in Section
\ref{sec-compare-mixed-state}.

\textit{We stress that the multiple-preparation approach requires 
what we call
``segmented
data''}, in the following sense:
to use an RCPS with the above procedure,
in the overall set of
$
\Ytwoqubitseqnb
\Ywritereadonestatenb
$
prepared states
defined above, 
one should know
which subset composed of
\ywritereadonestatenb\
prepared states
corresponds to a given state
value
\yketrandval,
in order
to estimate 
each
corresponding value
\yketbasisprobrandindexstdval\
as a sample frequency
over only that subset.
That segmentation is typically performed by
successively preparing the
\ywritereadonestatenb\
copies corresponding to the first drawn state
\yketrandval,
then the
\ywritereadonestatenb\
copies corresponding to the second drawn state,
and so on,
with a known value 
\ywritereadonestatenb.
The case of ``unsegmented data''
is discussed in Section
\ref{sec-compare-mixed-state}.

As stated above, from the point of view of
someone aiming at using an RCPS
(i.e. at reading it in our RWR procedure), the outcomes
\yqipout\ are considered to be randomly drawn.
How they are drawn, 
and therefore 
which
statistical
distributions 
are obtained for
these outcomes and 
for
the set of coefficients
$
\{
\Yketbasiscoefrandindexzeroval
,
\dots
,
\Yketbasiscoefrandindexlastval
\}
$%
,
depends on the considered
application.
For instance, in the above-mentioned communication
scenario, the receiver is the ``reader'' of our RWR
procedure, whereas the emitter is the ``writer'', who
prepares the states to be sent to the receiver.
The emitter may know the statistical distribution of the
states he prepares,
especially because the coefficients of the emitted
ket may be
defined by classical RV that may have known
statistical distributions.
Then, when the emitted ket is transferred through
the considered quantum channel to define the
received ket, the statistical distribution of the ket
coefficients is altered by that channel.
Similar considerations apply to the 
quantum parameter estimation
problem
discussed in Section
\ref{sec-appli},
where the method used for drawing the considered RV
is described.

The ket coefficients
\yketbasiscoefrandindexstd\
in
(\ref{eq-statepurerand})
may be expressed in polar form as
\begin{equation}
\Yketbasiscoefrandindexstd
=
\Yketbasiscoefrandindexstdmod
e^{
\Ysqrtminusone
\Yketbasiscoefrandindexstdphase
}
\label{eq-ketbasiscoefrandindexstd-polar}
\end{equation}
as also illustrated by the 
simplified 
single-qubit
form in
(\ref{eq-statepurerandonequbit}).
The measurements in the computational basis considered so far 
only
allow one to access (i.e. estimate)
the modulus parameters
\yketbasiscoefrandindexstdmod,
since
(\ref{eq-statepurerand-measprob})
yields
\begin{equation}
\Yketbasisprobrandindexstd
=
(
\Yketbasiscoefrandindexstdmod
)
^2
.
\label{eq-statepurerand-measprob-vs-ketbasiscoefrandindexstdmod}
\end{equation}
This also appears in
the 
simplified 
single-qubit
form in
(\ref{eq-statepurerand-measprob-onequbit-eq-one})-%
(\ref{eq-statepurerand-measprob-onequbit}).
Besides,  
measurements in bases other than the computational basis
(see p. 22 of
\ytextmodifartitionehundredsixtyfivevonestepone{\cite{booknielsen},}
and
\cite{amoi6-64})
provide 
information about
the phase parameters
\yketbasiscoefrandindexstdphase,
since one thus estimates the squared modulus of linear
combinations of the coefficients
\yketbasiscoefrandindexstd.
This e.g. corresponds to 
measuring 
$s_x$
spin components for
electron spins 1/2
whose overall state is expressed in the 
standard basis,
as detailed in 
\cite{amoi6-64}.
In 
the present
paper we only consider measurements in the 
computational basis, whereas other types of measurements for RCPS
will be
addressed in future papers.

The very general and major result obtained so far in this paper
is that \textit{the RCPS framework 
with
measurements in the computational basis
makes it possible to
access (estimates of) the
above-defined
probabilities
\yketbasisprobrandindexstd,
that are RV, and this then makes it
possible to exploit all their statistics,
e.g. to perform QIP tasks}.
The remainder of this paper shows
the wealth provided by
these statistics.
This will be especially
appreciated by contrasting the
capabilities thus reached
with those of the 
restricted approach to RCPS that is obtained by
using only the standard tools of quantum
mechanics.
Therefore, we first define 
that restricted
approach
hereafter.
\subsection{The density operator 
associated with an RCPS}
\label{sec-density-op-of-rcps}
In 
Chapter IV of
his famous book 
\cite{bookvonneumannfoundationsqm}, von Neumann first considers 
(deterministic-coefficient) pure 
states 
and claims
(p. 295):
``we succeeded in reducing all assertions of
quantum mechanics to the statistical formula
...'', where that formula 
defines the expectation (i.e. mean value)
of a physical quantity
\yphysicalquantitystdone\
and
reads
\yeqobservablemeanpurestate
with our notations, including
those defined in Section
\ref{sec-intro},
and
where
$E\{.\}$ stands for expectation,
here calculated for the considered state \yketdeterm.
Then considering mixed states (p. 296),
von Neumann further claims that
the density operator ``characterizes the mixture of states
just described completely,
with respect to its statistical properties'' and 
von Neumann then provides a formula 
that 
defines the expectation 
of 
\yphysicalquantitystdone\
for a mixed state
and
that here reads
\yeqobservablemeanmixedstate
with the above-defined notations.

Whereas the latter claim refers to the 
usual mixed states
\yopdensity\
considered by von Neumann,
one may wonder whether, for our RCPS too,
one only has to consider the 
mean of a physical quantity
\yphysicalquantitystdone\
and whether it can still be expressed as
$\Ytrace(\Yopdensity\Yobservableopstdone)$.
This leads to the preliminary question:
starting from and RCPS,
can one associate a density operator 
\yopdensity\
with it?
To this end, one should keep in mind that,
for a deterministic-coefficient pure state
(\ref{eq-statepuredeterm}),
we have
\yeqdetermcoefpurestatedensityop
so that the elements of the corresponding density matrix
read
\yeqdetermcoefpurestatedensitymatrixel
where
$^*$ stands for complex conjugate,
and
\yindexstdforketbasis\
and
\yindexstdtwoforketbasis\
range from 0 to
$
(
\Yketspacedim
-
1
)
$
as in
(\ref{eq-statepuredeterm}).
Therefore, as explained in
\cite{amoi6-67,amoi-arxiv-2022-adeville-artiti163v2}.
with an RCPS
defined by
(\ref{eq-statepurerand}),
one can 
associate a
density matrix whose
elements
read
\yeqdefdensoprandstateelindexstdindexstdtwo
In particular, its diagonal elements
read
\yeqdefdensoprandstateelindexstdindexstd
If
\yobservableopstdone\
is diagonal,
$\Ytrace(\Yopdensity\Yobservableopstdone)$
only depends on these diagonal elements
\ydensoprandstateelindexstdindexstd\
of
\yopdensity.

Eq.
(\ref{eq-defdensoprandstateelindexstdindexstd})
shows that the diagonal of the density matrix
only allows one to access very limited information
about the RV
\yketbasiscoefrandindexstd\
and
\yketbasisprobrandindexstd.
The quantity in
(\ref{eq-defdensoprandstateelindexstdindexstd})
may first be seen as 
a
\textit{second-order}
statistical parameter of
\yketbasiscoefrandindexstd,
whose classical counterpart 
is often called the ``mean power''
when considering its extension to a random signal
instead of an RV
\cite{book-papoulis,icabook-oja,bookkendallstuartvolone,amoi6-7,a247,a236}.
For a real-valued RV
\yketbasiscoefrandindexstd,
this parameter
$
E
\{
\Yketbasiscoefrandindexstd
^2
\}
$
is also 
the
second-order (non-centered) moment of this RV
(for a complex-valued 
\yketbasiscoefrandindexstd,
Eq. 
(\ref{eq-defdensoprandstateelindexstdindexstd})
therefore corresponds to
the second-order moment
of the RV
$
|
\Yketbasiscoefrandindexstd
|
$).
Eq.
(\ref{eq-defdensoprandstateelindexstdindexstd})
may also be seen as
the
\textit{first-order} moment
(i.e. expectation) of
\yketbasisprobrandindexstd.
The off-diagonal 
elements 
(\ref{eq-defdensoprandstateelindexstdindexstdtwo})
of
the density matrix
may yield additional information, but anyway
(i) this information is also limited to the
\textit{second-order}
statistics of the RV
\yketbasiscoefrandindexstd,
i.e. to a second-order joint moment 
which
is
their cross-correlation
and
(ii)
as mentioned above, this information cannot be accessed when
one only considers
$\Ytrace(\Yopdensity\Yobservableopstdone)$
and
\yobservableopstdone\
is diagonal.

In contrast, 
our approach, based on RCPS themselves,
yields
much richer information
because it allows one to access \textit{all the
statistics} of
\yketbasisprobrandindexstd,
as detailed further in this paper.
Besides, 
performing measurements in the computational basis for
a
\yketspacedim-dimensional RCPS
(\ref{eq-statepurerand})
yields estimates for
\yketspacedim\
RV
\yketbasisprobrandindexstd\
defined by
(\ref{eq-statepurerand-measprob}),
with
$0\leq\Yindexstdforketbasis\leq\Yketspacedim-1$.
Among these RV,
up to
$(\Yketspacedim - 1)$
may be statistically independent
because they sum to one, as shown by
(\ref{eq-statepurerandval-normalized}).
For
$\Yketspacedim > 2$,
one may therefore wonder whether
this set of
$(\Yketspacedim-1)>1$
quantities
provides richer information than the
\textit{single} scalar value
$\Ytrace(\Yopdensity\Yobservableopstdone)$
only considered in the usual approach.
This topic will  be investigated in future
papers
but,
in Section
\ref{sec-appli}, we show that, even for
$\Yketspacedim=2$,
our approach to RCPS based on the probabilities
\yketbasisprobrandindexstd\
is more powerful than the approach based on
the associated density operator.
\subsection{Exploiting 
higher-order statistics of RCPS}
\label{sec-exploit-hos}
As outlined above,
the statistics 
respectively accessible with
the RV
\yketbasisprobrandindexstd\
associated with an
RCPS and with
the approach based on
its density operator
\yopdensity\
and
$\Ytrace(\Yopdensity\Yobservableopstdone)$
yield a fundamental difference, which
will be better appreciated by first considering
the classical counterpart of this phenomenon.
Statistical methods for processing classical
random signals, images
or other types of data 
are often
limited
to the use of two types of parameters.
The first one is
their \textit{first-order statistics},
especially
the first-order moment, or expectation,
$E\{\boldsymbol{\mathbf{X}}\}$ of an RV
$\boldsymbol{\mathbf{X}}$.
The second one is
their \textit{second-order statistics},
which especially
include
(i) the second-order moment
$E\{\boldsymbol{\mathbf{X}}\boldsymbol{\mathbf{Y}}^*\}$ 
of RV
$\boldsymbol{\mathbf{X}}$
and
$\boldsymbol{\mathbf{Y}}$,
and
(ii)
the associated centered 
second-order moment,
i.e. covariance, 
$E\{\boldsymbol{\mathbf{\tilde{X}}}\boldsymbol{\mathbf{\tilde{Y}}}^*\}$ 
of
$\boldsymbol{\mathbf{X}}$
and
$\boldsymbol{\mathbf{Y}}$,
with the centered version of
$\boldsymbol{\mathbf{X}}$
defined as
$
\boldsymbol{\mathbf{\tilde{X}}}
=
\boldsymbol{\mathbf{X}}
-
E\{
\boldsymbol{\mathbf{X}}
\}
$
and the same for
$
\boldsymbol{\mathbf{\tilde{Y}}}
$.
Second-order statistics also include the restriction of
the above parameters to a single RV, i.e. when
$
\boldsymbol{\mathbf{X}}
=
\boldsymbol{\mathbf{Y}}
$,
which is connected with
mean power and variance, 
as
partly discussed above.

The above parameters were sufficient for
developing
powerful methods, such as Principal Component
Analysis (PCA)
\cite{book-jolliffe-pca,paper-abdi-williams-pca}
or Adaptive Noise Cancellation
(ANC)
\cite{a235,book-widrow-stearns}.
ANC
typically
makes it possible to restore an unknown signal
of interest from a measured signal that is a
so-called ``mixture'', i.e. combination,
of that useful signal and of noise,
but ANC requires
that another measurement provide the
noise signal alone.

In contrast, more difficult 
classical signal processing problems need
more advanced
tools, 
closely related to so-called
\textit{higher-order statistics} or HOS
(see e.g. the surveys in
\cite{a247,a236}
and more details in
\cite{icabook-oja,bookkendallstuartvolone,amoi6-7,book-comon-jutten-ap}).
``Higher'' here
means ``higher than 2''
and refers to the fact that these methods
(also) 
exploit other parts of the information contained in the
data than the above-defined first-order and second-order parameters.
In a basic form, this means exploiting $m$th-order moments
with
$m\geq3$,
these moments being defined as
$E\{\YqipRV^m\}$ 
for one real RV \yqipRV\
and
$E\{\YqipRV^{m_1}\YqipRVtwo^{m_2}\}$ 
with
$m_1+m_2=m$
for joint moments
of two real RV
\yqipRV\
and
\yqipRVtwo\
(and so on for more than two RV).
Here again, the corresponding \textit{centered}
moments are obtained by replacing
$\boldsymbol{\mathbf{X}}$
and
$\boldsymbol{\mathbf{Y}}$
by their centered versions
$
\boldsymbol{\mathbf{\tilde{X}}}
$
and
$
\boldsymbol{\mathbf{\tilde{Y}}}
$.
HOS methods also use (i) higher-order cumulants,
that may be expressed as specific combinations of
moments having attractive properties,
(ii) generalized moments
$E\{g(\YqipRV)\}$ 
and
$E\{g(\YqipRV)h(\YqipRVtwo)\}$
where $g$ and $h$ are 
arbitrary
nonlinear functions and
(iii) other quantities, that 
exploit all
the pdf
$f_{\YqipRV}$ or joint pdf
$f_{\YqipRV,\YqipRVtwo}$
of RV, 
such as differential entropy or mutual information
\cite{icabook-oja,bookkendallstuartvolone,amoi6-7,book-comon-jutten-ap,a247,a236}.
Besides, all these parameters extend to more than two RV,
as illustrated
below for their quantum version
(see
(\ref{eq-ketbasisprobrandindexstdjointmoment})).

In particular, the above tools have been used for classical
Independent Component Analysis (ICA) 
for so-called
i.i.d. signals.
ICA is a major class of methods for solving the
Blind Source Separation (BSS) 
well-known signal processing
problem, which consists of
extracting a set of source signals from measured signals that
\textit{all} are ``mixtures'', i.e. combinations, of these source signals.
ICA is a required extension of PCA and ANC
because, for i.i.d. signals,
the above-defined BSS problem cannot be solved with only second-order 
statistical methods,
including PCA and ANC,
but it can be solved by exploiting the additional information 
that is provided
by HOS for
non-Gaussian signals
and that is used in ICA
(see details e.g. in
Chapter 7 of
\cite{icabook-oja}
or in
Chapter 12 of
\cite{%
amoi6-7}%
).
This problem is also closely related to blind system or
mixture identification
\cite{a593,amoi6-48,amoi6-144},
because BSS and hence ICA essentially require one to
estimate the inverse of the function, i.e. ``system'',
that mixes the source signals.

Having the above classical data processing background in mind,
we now move to the quantum framework. Our
approach based on the RV
\yketbasisprobrandindexstd\
associated with an RCPS may then be expected 
to be able to solve QIP
problems that cannot be handled by restricting oneself to 
(i) the
density operator
\yopdensity\
associated with an RCPS
and (ii) the corresponding
mean of observable
$\Ytrace(\Yopdensity\Yobservableopstdone)$.
More precisely,
to extend QIP capabilities, one may exploit the HOS of the RV 
\yketbasiscoefrandindexstd\
through
the 
statistics of the RV
\yketbasisprobrandindexstd\
at orders higher than one
whereas, as explained above,
\yopdensity\
and
$\Ytrace(\Yopdensity\Yobservableopstdone)$
essentially
access the first-order statistics of
\yketbasisprobrandindexstd\
and anyway only the second-order statistics
of \yketbasiscoefrandindexstd.

In a basic form, this means exploiting fourth-order parameters
of any 
\yketbasiscoefrandindexstd\
with
$0\leq\Yindexstdforketbasis\leq\Yketspacedim-1$,
through
the second-order moment, 
i.e. mean power,
$
E \{
(
\Yketbasisprobrandindexstd
)
^2
\}
$
of
\yketbasisprobrandindexstd,
or through
is centered second-order moment, 
i.e. variance,
\begin{equation}
E
\{
(
\Yketbasisprobrandindexstd
-
E \{
\Yketbasisprobrandindexstd
\}
)
^2
\}
=
E \{
(
\Yketbasisprobrandindexstd
)
^2
\}
-
(
E \{
\Yketbasisprobrandindexstd
\}
)
^2
.
\label{eq-variance}
\end{equation}
Other statistical parameters 
of the RV
\yketbasisprobrandindexstd\
may also be considered by
extending, to the quantum framework, the parameters that
we summarized above for the classical framework.
This first includes parameters for a single RV
\yketbasisprobrandindexstd,
such as various higher-order moments 
$E\{
(
\Yketbasisprobrandindexstd
)
^m
\}$ 
or generalized
moments
$
E \{
g
(
\Yketbasisprobrandindexstd
)
\}
$.
Importantly, this also includes parameters
associated with \textit{several} of these
RV, such as their joint moments
\yeqketbasisprobrandindexstdjointmoment
where
$
I
$
is an arbitrary subset of the set of
indices
\yindexstdforketbasis\
with
$0\leq\Yindexstdforketbasis\leq\Yketspacedim-1$
and
$
m
_{
\Yindexstdforketbasis
}
$
are integers, that define the overall order of the
considered moment
\footnote{The complete class
of moments that may be 
introduced 
for a
given set of ket coefficients \yketbasiscoefrandindexstd\ is
defined as the expectation of an arbitrary product of factors,
where each factor is freely selected to be
either 
a coefficient \yketbasiscoefrandindexstd\
or its conjugate
(and both forms may appear for any given
\yketbasiscoefrandindexstd). 
Here, we only consider a subset of these moments.
This is due to the fact that
we start from the probabilities
\yketbasisprobrandindexstd\
(because they are the quantities that we access with measurements),
not the coefficients
\yketbasiscoefrandindexstd,
and we then build the quantities
(\ref{eq-ketbasisprobrandindexstdjointmoment}).
If expressing these quantities with respect to ket coefficients,
every factor
\yketbasiscoefrandindexstd\
in
(\ref{eq-ketbasisprobrandindexstdjointmoment})
is constrained to appear
together with its conjugate, because
(\ref{eq-statepurerand-measprob})
shows that
\yketbasisprobrandindexstd\
is the product of
\yketbasiscoefrandindexstd\
and its conjugate.%
}.

How the above statistical parameters
are used depends on the considered
QIP task. A large set of potential applications deal
with the estimation of parameters of a quantum system
(or of a quantum state),
therefore with a close relationship with
quantum process (or state) tomography
and with Hamiltonian 
estimation.
The resulting 
classes of
QIP 
methods especially include
the quantum extension
of so-called
moment matching methods used for classical data
processing (see e.g. Section 4.3 of
\cite{icabook-oja}).
This consists of expressing the RV 
\yketbasisprobrandindexstd,
and then some of their moments,
with
respect to quantities
including the unknown parameters (e.g. of the considered
system) to be estimated.
Each such
moment thus
yields an equation with respect to the
unknown 
parameters.
Estimates of these moments may be derived
from measurements as explained above.
One then
uses these estimates instead of the actual
moments in the above equations.
Considering enough moments thus yields enough
equations,
from which the values of the
unknown
parameters are derived.
These values are therefore those that match
the estimated moments, hence the name of this
approach.

Although we did not explicitly mention that
quantum
moment matching concept
in our 
application-driven 
QIP
papers,
we already used it 
in several of them%
: 
see 
e.g.
\cite{amoi5-31,amoi6-18,amoi-ieee-tqe-2021},
\cite{amoi6-42} (Section 1.7.2),
\cite{amoi6-118}.
These applications concerned the quantum version of
BSS, i.e. BQSS, and of blind system and parameter identification, i.e. BQPT
and BHPE.
They were focused on a specific type of 
quantum process/system
(which corresponds to the mixing function of
classical BSS):
we considered two qubits coupled according to the Heisenberg model.
In addition,
a 
new 
application of the above quantum moment matching
procedure is detailed below in Section
\ref{sec-appli}.
This new investigation
has complementary features with respect to our above-mentioned
previous works.
First,
whereas we previously only considered 
the statistics of the probabilities
\yketbasisprobrandindexstd\
associated with an
RCPS, we here moreover
compare the capabilities
thus achieved with those of the appproach
to 
RCPS
based on
\yopdensity\
and
$\Ytrace(\Yopdensity\Yobservableopstdone)$
that we defined in Section
\ref{sec-density-op-of-rcps}.
We
thus explicitly prove that
the approach based on the statistics of
\yketbasisprobrandindexstd\
is more powerful.
Besides,
we
only used first-order moments
of
\yketbasisprobrandindexstd\
in the 
previous works
\cite{amoi5-31,amoi6-18,amoi-ieee-tqe-2021},
\cite{amoi6-42} (Section 1.7.2),
\cite{amoi6-118}
(we also used other statistical parameters, but for
quantities that are only indirectly related to
\yketbasisprobrandindexstd:
see
\footnote{Our previous investigations related to BQSS, BQPT and BHPE
involve a quantum process.
The above-mentioned papers
\cite{amoi5-31,amoi6-18,amoi-ieee-tqe-2021},
\cite{amoi6-42} (Section 1.7.2),
\cite{amoi6-118}
directly use statistical parameters of the probabilities
\yketbasisprobrandindexstd\
of measurements performed at the output of that process.
In contrast, other investigations, dealing with BQSS,
first use the individual values of these classical-form data
\yketbasisprobrandindexstd\
as the input of a classical processing system, called the separating
system.
The outputs of that system aim at restoring 
modulus parameters and combinations of phase parameters
of coefficients of several single-qubit quantum states.
These parameters thus have some relationships 
with
the quantities
\yketbasiscoefrandindexstdmod\
and
\yketbasiscoefrandindexstdphase\
in
(\ref{eq-ketbasiscoefrandindexstd-polar}).
These BQSS methods are based on various statistical 
parameters of the outputs of the separating
system:
their generalized moments
are used in
Section 1.7.3 of
\cite{amoi6-42}
and their cumulants 
in
\cite{amoi6-9},
whereas their whole
pdf are exploited through 
their mutual information
(see Section 1.5 of
\cite{amoi6-42}),
with a connection with the maximum likelihood approach
(see Section 1.6 of
\cite{amoi6-42}).
All these approaches thus have an indirect link with the 
higher-order statistics of the
\yketbasiscoefrandindexstdmod\
and
\yketbasiscoefrandindexstdphase\
parameters, and hence with those of
\yketbasisprobrandindexstd.}).
In constrast,
we here also take advantage 
of the second-order moment
of
\yketbasisprobrandindexstd.
Moreover,
we here 
investigate
the estimation 
of parameters of a quantum
process, which is a task closely related to BQPT and BHPE,
but we here consider 
a different class of
processes.
That class
is much more general than 
the above-mentioned
Heisenberg process
in the sense that
it addresses
{any
energy-preserving process}, represented by an arbitrary
unitary matrix, although it is
only considered for a single qubit for the sake of clarity.
Finally, we not only propose blind estimation methods,
but also non-blind ones.

Before we focus on that QIP task in Section
\ref{sec-appli},
the remainder of the present section is dedicated to the
presentation of other general features
of RCPS.
We 
first
stress that a 
very
large number of moments
(\ref{eq-ketbasisprobrandindexstdjointmoment}),
and therefore e.g.
of moment-based equations
in the above quantum moment matching procedure,
may be defined from the same set of
measurements.
This is very attractive because it may
drastically reduce
the 
number of types of measurements
required to estimate the parameters of interest,
whereas this
is 
currently
a bottleneck as soon as the dimensionality
of the considered system or state increases.
The experimental complexity 
of performing various types of quantum measurements
(e.g. spin components along various directions)
will thus be, at least partly, replaced by additional processing
of a reduced set of
measurement results on a classical computer, which is much 
simpler.
More precisely,
various
previously reported QIP methods 
use only a \textit{single}
quantity, namely
the first-order moment
(i.e.
the mean),
for each type of measurement, 
and they therefore
require various types of measurements to obtain enough information
about the considered phenomenon.
In contrast, our approach based on RCPS
can
get enough information with a lower number of 
types of measurement, by exploiting \textit{various parameters} of
the quantities 
\yketbasisprobrandindexstd,
including
their mean power and
higher-order statistical parameters,
derived on a classical computer from all measurement results
obtained for each given type of measurement.
We plan to investigate this topic in future papers
for general configurations, but we already illustrate it with an example
in Section
\ref{sec-appli}
of the present paper.
\subsection{Limitations 
of 
usual
statistics 
of observables}
\label{sec-limit-observable}
We stress that, for a given physical quantity
\yphysicalquantitystdone\
and a given RCPS
\yketrand,
\textit{%
the approach proposed in this paper exploits
statistical parameters
of 
(one or several)
RV
\yketbasisprobrandindexstd,
not those of the 
(single)
RV defined by the measured values of
\yphysicalquantitystdone}
(note also that the RV
\yketbasisprobrandindexstd\
may be continuous-valued or discrete-valued as discussed
in Section
\ref{sec-compare-mixed-state}, whereas the RV defined by
\yphysicalquantitystdone\
is 
generally
discrete-valued).
Our motivation is that 
this approach based on 
the statistical parameters
of interest
of
\yketbasisprobrandindexstd\
yields much more information
than the 
usual 
approach based on
\yphysicalquantitystdone,
as will now be shown.
To this end, we hereafter
first revisit the concept of the mean of
a physical quantity,
that we only partly addressed is Section
\ref{sec-density-op-of-rcps},
but now without resorting to the density operator of an RCPS.
This then allows us to naturally
proceed further, by combining the approach
used here with some HOS concepts introduced in Section
\ref{sec-exploit-hos}.

Let us first consider the mean, 
hence the
first-order statistics,
of
\yphysicalquantitystdone.
Using an arbitrary orthonormal
basis
$\{\Yketbasisindexstd\}$,
the RCPS
\yketrand\
is defined by
(\ref{eq-statepurerand}),
whereas
\yphysicalquantitystdone\
is represented by a possibly non-diagonal matrix
whose elements are denoted as
\yobservableopstdoneelindexstdoneindexstdtwo.
The usual expression 
(\ref{eq-observablemeanpurestate})
of the mean of
\yphysicalquantitystdone\
for a deterministic-coefficient pure state is here first used for
the state
\yketrandval\
associated with a single outcome
\yqipout.
This
yields
\yeqphysicalquantitystdonemeanforketrandval
Then using the expectation of the latter quantity 
over all outcomes
\yqipout\
yields the
mean of
\yphysicalquantitystdone\
for
the RCPS
\yketrand,
which reads
\yeqphysicalquantitystdonemeanforketrand

That mean of
\yphysicalquantitystdone\
therefore has two 
limitations.
First, it is only 
related to 
(part of)
the \textit{second-order} statistics
of the RV
\yketbasiscoefrandindexstd,
that include two aspects:
\begin{enumerate}
\item
Moments
that each involve
a single RV.
They
correspond to
the terms
with
$
\Yobservableopstdoneindexelone
=
\Yobservableopstdoneindexeltwo
$
in
(\ref{eq-physicalquantitystdonemeanforketrand}),
namely to the 
probabilities
\yketbasisprobrandindexstd\
defined by
(\ref{eq-statepurerand-measprob}).
\item
Joint moments of
two RV, that correspond to
the terms
with
$
\Yobservableopstdoneindexelone
\neq
\Yobservableopstdoneindexeltwo
$
in
(\ref{eq-physicalquantitystdonemeanforketrand}).
When
\yphysicalquantitystdone\
is represented by a diagonal matrix,
these terms
disappear from
(\ref{eq-physicalquantitystdonemeanforketrand}).
\end{enumerate}

When the dimension
\yketspacedim\
of the state space is higher than 2,
using only the mean of
\yphysicalquantitystdone\
yields an additional limitation:
estimating that mean yields only a \textit{single}
equation with respect to estimates of 
(some: see above) 
statistics 
of all
RV
$
\boldsymbol{\mathbf{
\Yketbasiscoefdetermnot
_{
\Yobservableopstdoneindexelone
}
}}
^*
\boldsymbol{\mathbf{
\Yketbasiscoefdetermnot
_{
\Yobservableopstdoneindexeltwo
}
}}
$,
including all
\yketbasisprobrandindexstd,
as shown by
(\ref{eq-physicalquantitystdonemeanforketrand}).
In
contrast, our approach 
based on the
probabilities
\yketbasisprobrandindexstd\
of an RCPS
themselves
allows one
to separately estimate 
(all) 
the
statistics of
\textit{each} 
of these probabilities.
In the specific case when
$
\Yketspacedim
=
2
$,
i.e. for a single qubit,
this difference between the considered two
approaches 
reduces,
because only one
independent probability
\yketbasisprobrandindexstd\
exists, as shown by
(\ref{eq-statepurerand-measprob-onequbit}),
but several statistical parameters of that
\yketbasisprobrandindexstd\
can still be exploited, as explained above
\footnote{Although the mean of an observable is here 
intentionally analyzed
without resorting to the content of Section
\ref{sec-density-op-of-rcps},
these two parts of this paper are clearly connected, because
(\ref{eq-physicalquantitystdonemeanforketrand})
is nothing but the quantity
$
\Ytrace(\Yopdensity
$%
\^A)
defined in Section
\ref{sec-density-op-of-rcps} for an RCPS,
and the discussion provided after
(\ref{eq-physicalquantitystdonemeanforketrand})
therefore has connections with the comments we made in
Section
\ref{sec-density-op-of-rcps},
mainly about the density operator 
\yopdensity\
of an RCPS
and partly about the resulting
$
\Ytrace(\Yopdensity
$%
\^A).}.

One may then try to access richer information
by considering the mean 
$
E
\{
\Yqipfuncstdone
(
\Yphysicalquantitystdone
)
\}
_{\Yketrand}
$
of \textit{a function
\yqipfuncstdone\
of
\yphysicalquantitystdone},
as the quantum counterpart of the generalized
moments 
$E\{g(\YqipRV)\}$ 
of classical RV,
and similarly to the quantum
generalized
moments
$
E \{
g
(
\Yketbasisprobrandindexstd
)
\}
$,
both
defined in Section
\ref{sec-exploit-hos}.
Here,
\yqipfuncstdone\
is an arbitrary function, 
and this e.g. includes the 
specific case
when
\begin{equation}
\Yqipfuncstdone
(
x
)
=
\left(
x
-
E
\{
\Yphysicalquantitystdone
\}
_{\Yketrand}
\right)
^2
\end{equation}
for which
$
E
\{
\Yqipfuncstdone
(
\Yphysicalquantitystdone
)
\}
_{\Yketrand}
$
is the variance of
\yphysicalquantitystdone\
for the RCPS
\yketrand\
(this is coherent with the corresponding expression of the variance
for a usual, i.e. deterministic-coefficient, pure state:
see e.g. p. 295 of
\cite{bookvonneumannfoundationsqm}).
However, even
for arbitrary 
functions
\yqipfuncstdone,
that approach 
based on
$
E
\{
\Yqipfuncstdone
(
\Yphysicalquantitystdone
)
\}
_{\Yketrand}
$
has limited capabilities, as will now be shown.
Using an arbitrary orthonormal
basis
$\{\Yketbasisindexstd\}$,
the expression of the matrix
that represents
$
\Yqipfuncstdone
(
\Yphysicalquantitystdone
)
$
may be derived from the considered
physical quantity
\yphysicalquantitystdone\
and function
\yqipfuncstdone:
see
e.g.
\cite{livreperes1995}.
Its elements are hereafter denoted as
$
g
_{
\Yobservableopstdoneindexelone
\Yobservableopstdoneindexeltwo
}
$
and their expressions are not needed here:
using the same
approach as in
(\ref{eq-physicalquantitystdonemeanforketrandval})-%
(\ref{eq-physicalquantitystdonemeanforketrand})
yields
\yeqphysicalquantitystdonefuncstdonemeanforketrandversiontwo
again with the connection
(\ref{eq-statepurerand-measprob})
with the probabilities
\yketbasisprobrandindexstd\
for the terms
of
(\ref{eq-physicalquantitystdonefuncstdonemeanforketrandversiontwo})
with
$
\Yobservableopstdoneindexelone
=
\Yobservableopstdoneindexeltwo
$.
The main conclusion and limitation that may be derived from
(\ref{eq-physicalquantitystdonefuncstdonemeanforketrandversiontwo})
is that this quantity too only depends on the
\textit{second-order} statistics of the coefficients
$
\boldsymbol{\mathbf{
\Yketbasiscoefdetermnot
_{
\Yobservableopstdoneindexelone
}
}}
$:
introducing the function
\yqipfuncstdone\
yields a nonlinearity in the expressions of the
matrix elements
$
g
_{
\Yobservableopstdoneindexelone
\Yobservableopstdoneindexeltwo
}
$
\cite{livreperes1995},
not in the statistics of the coefficients
$
\boldsymbol{\mathbf{
\Yketbasiscoefdetermnot
_{
\Yobservableopstdoneindexelone
}
}}
$
\footnote{It should however be noted that using
the above
function
\yqipfuncstdone\
has a possibly attractive effect:
(\ref{eq-physicalquantitystdonefuncstdonemeanforketrandversiontwo})
allows one to access a different
linear combination of the 
probabilities
\yketbasisprobrandindexstd\
(and cross-terms
$
E
\{
\boldsymbol{\mathbf{
\Yketbasiscoefdetermnot
_{
\Yobservableopstdoneindexelone
}
}}
^*
\boldsymbol{\mathbf{
\Yketbasiscoefdetermnot
_{
\Yobservableopstdoneindexeltwo
}
}}
\}
$)
than
(\ref{eq-physicalquantitystdonemeanforketrand}).
Jointly considering
(\ref{eq-physicalquantitystdonefuncstdonemeanforketrandversiontwo})
for various functions
\yqipfuncstdone\
and solving the corresponding equations
might therefore provide a way to separately estimate
the expectation of each 
probability
\yketbasisprobrandindexstd.
Anyway, it then remains that:
1)
our approach directly based on these
probabilities
\yketbasisprobrandindexstd\
also makes it possible to estimate
their expectations and without having to
create and solve the above equations,
and 2)
the approach based on
the mean of observables and of function of observables
only accesses these \textit{expectations} of
\yketbasisprobrandindexstd\
(and the other \textit{second-order} parameters
$
E
\{
\boldsymbol{\mathbf{
\Yketbasiscoefdetermnot
_{
\Yobservableopstdoneindexelone
}
}}
^*
\boldsymbol{\mathbf{
\Yketbasiscoefdetermnot
_{
\Yobservableopstdoneindexeltwo
}
}}
\}
$
of the 
ket coefficients),
not their other statistics, unlike our approach.
Our approach directly based on (all the statistics of)
the probabilities
\yketbasisprobrandindexstd\
therefore remains of much higher
interest.},
\footnote{The mean of \textit{a function of} an observable was not
explicitly addressed in Section
\ref{sec-density-op-of-rcps}
and was therefore independently detailed in the present section.
However, its connection with Section
\ref{sec-density-op-of-rcps}
may be shown as follows.
$
G
=
\Yqipfuncstdone
(
\Yphysicalquantitystdone
)
$
is nothing but another observable,  with an associated operator
\^G.
Eq. 
(\ref{eq-physicalquantitystdonefuncstdonemeanforketrandversiontwo})
defines the mean 
$
E
\{
G
\}
_{\Yketrand}
$
of that new observable, that could also be expressed
as
$
\Ytrace(\Yopdensity
$%
\^G)
and that therefore has the limitations
that we defined 
for
$
E
\{
\Yphysicalquantitystdone
\}
_{\Yketrand}
=
\Ytrace(\Yopdensity
$%
\^A)
in
Section
\ref{sec-density-op-of-rcps}
and at the beginning of the present section,
when considering an \textit{arbitrary} observable
\yphysicalquantitystdone.}.
\subsection{Other connection of RCPS with density operators}
Another
connection between RCPS and 
the usual framework of 
deterministic-coefficient pure states 
is now introduced as follows.
Starting from an RCPS
\yketrand,
we consider each associated
deterministic-coefficient pure state
\yketrandval.
We
use
its density operator in the
usual sense of quantum mechanics: it is defined by adapting
(\ref{eq-determcoefpurestatedensityop})
and
(\ref{eq-determcoefpurestatedensitymatrixel})
to
\yketrandval\
instead
of
\yketdeterm.
This yields
\yeqrandstatedensitymatrixelapproachtwoout
where we
denote as
\ydensoprandstateapproachtwoout\
the density matrix and density operator of
\yketrandval.
We moreover introduce the original
\textit{random} operator and the associated
random
matrix, both denoted as
\ydensoprandstateapproachtwo,
as follows:
it is the
operator/matrix which depends on the outcome
\yqipout\
and
whose 
realization
associated with any outcome
\yqipout\
is
\ydensoprandstateapproachtwoout.
The elements of the matrix
\ydensoprandstateapproachtwo\
then read
\yeqrandstatedensitymatrixelapproachtworv

This random operator
\ydensoprandstateapproachtwo\
thus consists of an ensemble
of usual density operators
\ydensoprandstateapproachtwoout.
It
should be distinguished
from the single, deterministic,
density
operator
\yopdensity\
defined by
(\ref{eq-defdensoprandstateelindexstdindexstdtwo}),
that we 
previously
associated with an RCPS.
Yet, they are closely connected, since
\yopdensity\
is the expectation of
\ydensoprandstateapproachtwo,
as shown by
(\ref{eq-defdensoprandstateelindexstdindexstdtwo})
and
(\ref{eq-randstatedensitymatrixelapproachtwoorv}).
Besides,
(\ref{eq-randstatedensitymatrixelapproachtwoorv})
shows that
the diagonal elements of
\ydensoprandstateapproachtwo\
are nothing but the quantities
\yketbasisprobrandindexstd\
that we previously introduced in
(\ref{eq-statepurerand-measprob}).
This operator
\ydensoprandstateapproachtwo\
therefore also contains the wealth of 
all the statistics of
the random probabilities
\yketbasisprobrandindexstd\
upon which we focus in this paper,
plus its off-diagonal elements 
to be further investigated.
The random operator
\ydensoprandstateapproachtwo\
associated with the above-mentioned ensemble 
of
\ydensoprandstateapproachtwoout\
is thus much
richer than its plain expectation consisting of the
density operator
\yopdensity\
of
(\ref{eq-defdensoprandstateelindexstdindexstdtwo}).

Besides, it is thus not surprising that we succeeded in
associating \textit{several} RCPS 
(in the sense of
(\ref{eq-defdensoprandstateelindexstdindexstdtwo}))
with a given density operator
in our very recent investigation
\cite{amoi-arxiv-2022-adeville-artiti163v2}:
knowing the mean
operator
\yopdensity\
is not sufficient for 
imposing all the statistics
of the coefficients
\yketbasiscoefrandindexstd\
of an RCPS 
nor
those of
its
random
operator
\ydensoprandstateapproachtwo\
(similarly, 
knowing the mean of a classical RV
is not sufficient for 
imposing all the statistics
of that RV).

If one would like to use
all
\ydensoprandstateapproachtwo,
one would then have to define how to access related properties in
practice, typically by means of measurements, 
as we did above for
\yketbasisprobrandindexstd,
i.e. for the diagonal of
\ydensoprandstateapproachtwo.
The non-diagonal elements of
\ydensoprandstateapproachtwo\
will be analyzed in our future papers, 
whereas we keep on focusing
on
\yketbasisprobrandindexstd\
hereafter.
\section{%
An
application to
quantum parameter estimation}
\label{sec-appli}
\subsection{Considered quantum system and task}
\label{sec-appli-system}
A well-known QIP task is Quantum Process
Tomography (QPT),
especially
\footnote{%
See also
\cite{booknielsen}
p. 398 for the other earliest references.%
}
introduced in 1997 in 
\cite{amq30official}%
.
QPT
is the quantum
version
of classical 
non-blind
system identification
(see
e.g.
\cite{booknielsen,
amq-baldwin-physreva-2014,
amq75,
amq45official,
amq50-physical-review,
amq59,
amq48,
amq52-physical-review,
amq56,
amq41}%
)
and
is also closely connected 
with
non-blind
quantum channel estimation and 
phase estimation
\cite{amoi-ieee-tqe-2021}.
It e.g. applies to
a quantum system 
that 
here
does not
interact with 
its environment,
whose input
is 
here 
an RCPS
\yketrandprocessin\
equal to the initial state of the system,
and whose output is then 
an RCPS
\yketrandprocessout\
equal to the final state of the system.
The process/transform
applied by the system to its input
is unknown and is to be
identified, i.e. estimated.
It is represented by a unitary matrix
\yopmix:
multiplying the 
vector of coefficients of the
input ket 
\yketrandprocessin\
by that matrix
yields the 
vector of coefficients of the
output ket
\yketrandprocessout\
(see
(\ref{eq-vector-coef-out-vs-in-onequbit})
below for an example).

For a given initial-to-final time interval,
the expression of the above matrix
\yopmix\
is defined by the Hamiltonian of the
quantum system, which may be known to belong
to a given class, whereas the values of the
parameters of that model are unknown and are
to be estimated. A related task is therefore
(non-blind)
Hamiltonian Parameter Estimation (HPE)
\cite{amq99,
amq103,
amq102}.
Such parameter estimation problems
are also addressed, but often referred to as Hamiltonian identification,
e.g. in
\cite{amq97,
amq56,
amq101}
and partly
\cite{amq100}.

Standard QPT and HPE methods are non-blind in the
sense that they estimate the considered
quantities by knowing 
the input values of
the process, in addition to 
measurement results associated with
its output.
We extended these approaches to their blind
version, which is more powerful because it
does not require one to know each value of the
applied input but only some of their
statistical properties:
see e.g. our previous works in
\cite{amoi6-46,amoi6-79,amoi6-118}
for blind QPT (BQPT) 
and
\cite{amoi-ieee-tqe-2021}
for blind HPE (BHPE).

As stated above,
these previous investigations of blind methods
were focused on
a specific class of 
two-qubit processes and associated Hamiltonian,
based on cylindrical-symmetry Heisenberg coupling.
In contrast, we here consider 
a very generic 
class of processes:
we address any unitary process, 
yet focusing
on single-qubit processes.
Single-qubit
processes are considered
both for the
sake of clarity 
and to show that our
approach to RCPS 
based
on
probability statistics
yields better performance
than the approach to RCPS based on the density
operator 
and
$\Ytrace(\Yopdensity\Yobservableopstdone)$
even for a single qubit, i.e. when
the wealth of our approach does not result from
the availability of \textit{several} independent probabilities
\yketbasisprobrandindexstd\
(see Section 
\ref{sec-density-op-of-rcps}).

A model representing all single-qubit unitary
processes 
is obtained by 
expressing the above matrix
\yopmix\
as follows
(see
\cite{booknielsen}
p. 176):
\begin{equation}
\Yopmix
=
e
^{
\Ysqrtminusone
\Yopmixparamone
}
\left[
\begin{tabular}{ll}
$
e
^{
\Ysqrtminusone
(
-
\Yopmixparamtwo
/
2
-
\Yopmixparamfour
/
2
)
}
\cos
\left(
\frac{\Yopmixparamthree}{2}
\right)
$
&
$
-
e
^{
\Ysqrtminusone
(
-
\Yopmixparamtwo
/
2
+
\Yopmixparamfour
/
2
)
}
\sin
\left(
\frac{\Yopmixparamthree}{2}
\right)
$
\\
$
e
^{
\Ysqrtminusone
(
\Yopmixparamtwo
/
2
-
\Yopmixparamfour
/
2
)
}
\sin
\left(
\frac{\Yopmixparamthree}{2}
\right)
$
&
$
e
^{
\Ysqrtminusone
(
\Yopmixparamtwo
/
2
+
\Yopmixparamfour
/
2
)
}
\cos
\left(
\frac{\Yopmixparamthree}{2}
\right)
$
\end{tabular}
\right]
.
\label{eq-def-opmix-qubitone-unitary}
\end{equation}

The problem addressed below is the estimation of
all or at least part 
of the parameters
\yopmixparamone\
to
\yopmixparamfour.
To this end,
we propose 
both non-blind and blind estimation methods.

Since the output state of the considered process and
hence the matrix
\yopmix\
are defined only up to a phase factor, one may
anticipate that
\yopmixparamone\
cannot be estimated
(and that this is not an issue).
This is
confirmed by the operation of the
methods proposed below.
\subsection{Considered states and measurements}
\label{sec-appli-state-and-meas}
The random-coefficient state 
\yketrandprocessin\
applied 
to
the input of the
considered process is defined by
the right-hand term of
(\ref{eq-statepurerandonequbit}).
The resulting output state 
\yketrandprocessout\
of that
process
is defined by
the right-hand term of
(\ref{eq-statepurerand})
with
$
\Yketspacedim
=
2
$.
Its
coefficients
\yketbasiscoefrandindexstd\
here form the vector
\begin{equation}
\left[
\begin{tabular}{l}
\yketbasiscoefrandindexzero
\\
\yketbasiscoefrandindexone
\end{tabular}
\right]
=
\Yopmix
\left[
\begin{tabular}{l}
\Yketbasiscoefrandindexzeroamongtwomod
\\
$
\sqrt{
1
-
\Yketbasiscoefrandindexzeroamongtwomod
^2
}
e^{
\Ysqrtminusone
\Yketbasiscoefrandindexoneamongtwophase
}
$
\end{tabular}
\right]
.
\label{eq-vector-coef-out-vs-in-onequbit}
\end{equation}

Measurements are then performed for copies of 
each realization of
the state
\yketrandprocessout\
that corresponds to an outcome
\yqipout.
In a practical QIP setup,
only some types
of measurements are 
allowed.
To perform a fair comparison of the two processing methods
respectively
based on the probabilities
\yketbasisprobrandindexstd\
and on the density operator
\yopdensity,
both methods should be considered for the same
type(s) of measurements.
We hereafter
analyze
the case when only measurements in the
computational basis are allowed
\footnote{%
\label{footnote-other-measurements}
One may expect that higher performance can
be
obtained by also considering other types
of measurements, but this is true for both
methods and our goal here is not to derive their
ultimate performance depending on the considered
measurements
but to compare their capabilities for a given,
relevant, type of measurements}.
From a physical point of view, this e.g. corresponds
to implementing the considered qubit as a
spin 1/2 and measuring its
$s_z$ spin component
(the basis vectors 
$| 0 \rangle$
and
$| 1 \rangle$
in
(\ref{eq-statepurerandonequbit})
might then be denoted as
$| + \rangle$
and
$| - \rangle$).
These measurements have two possible results, whose
probabilities are
defined by
(\ref{eq-statepurerand-measprob}).
Using
(\ref{eq-def-opmix-qubitone-unitary})
and
(\ref{eq-vector-coef-out-vs-in-onequbit}),
this may be shown to
yield
\begin{eqnarray}
\Yketbasisprobrandindexzero
&
=
&
\cos
(
\Yopmixparamthree
)
\Yketbasiscoefrandindexzeroamongtwomod
^2
+
\frac{
1
-
\cos
(
\Yopmixparamthree
)
}
{2}
\nonumber
\\
&
&
-
\cos
(
\Yopmixparamfour
+
\Yketbasiscoefrandindexoneamongtwophase
)
\sin
(
\Yopmixparamthree
)
\Yketbasiscoefrandindexzeroamongtwomod
\sqrt{
1
-
\Yketbasiscoefrandindexzeroamongtwomod
^2
}
\label{eq-qubitone-unitary-processout-measprob-first}
\\
\Yketbasisprobrandindexone
&
=
&
1
-
\Yketbasisprobrandindexzero
.
\label{eq-qubitone-unitary-processout-measprob-second}
\end{eqnarray}
\subsection{Approach based on the 
mean value
$\Ytrace(\Yopdensity\Yobservableopstdone)$}
\label{sec-appli-method-mean-value}
We first investigate 
an approach 
based on the
principles presented
in Section
\ref{sec-density-op-of-rcps}.
In the considered basis,
the measured physical quantity, as defined in Section
\ref{sec-appli-state-and-meas},
is represented by the matrix
\begin{equation}
\Yobservableopstdone
=
\left[
\begin{tabular}{cc}
$
\frac{1}{2}
$
&
0
\\
0
&
$
-
\frac{1}{2}
$
\end{tabular}
\right]
.
\end{equation}
Therefore
\begin{equation}
\Ytrace(\Yopdensity\Yobservableopstdone)
=
\frac{1}{2}
(
\Ydensoprandstate
_{
00
}
-
\Ydensoprandstate
_{
11
}
)
.
\end{equation}
Using
(\ref{eq-defdensoprandstateelindexstdindexstd})
and
(\ref{eq-qubitone-unitary-processout-measprob-second}),
this yields
\begin{equation}
\Ytrace(\Yopdensity\Yobservableopstdone)
=
E
\{
\Yketbasisprobrandindexzero
\}
-
\frac{1}{2}
.
\label{eq-qubitone-unitary-trace-vs-prob}
\end{equation}
As an example,
for all
quantum parameter estimation methods investigated
in this paper,
we moreover set the 
same 
following
constraints on the statistics 
of the input state
\yketrandprocessin\
(but not on its 
individual
values).
\yketbasiscoefrandindexzeroamongtwomod\
and
\yketbasiscoefrandindexoneamongtwophase\
are statistically independent RV.
\yketbasiscoefrandindexzeroamongtwomod\
has a uniform distribution over the interval
$
[
\Yketbasiscoefrandindexzeroamongtwomoddistribboundlow
,
\Yketbasiscoefrandindexzeroamongtwomoddistribboundhigh
]
$
and
\yketbasiscoefrandindexoneamongtwophase\
has a uniform distribution over the interval
$
[
-
\Yketbasiscoefrandindexoneamongtwophasedistribbound
,
\Yketbasiscoefrandindexoneamongtwophasedistribbound
]
$,
where
\yketbasiscoefrandindexzeroamongtwomoddistribboundlow,
\yketbasiscoefrandindexzeroamongtwomoddistribboundhigh\
and
\yketbasiscoefrandindexoneamongtwophasedistribbound\
are free parameters.
In these conditions,
(\ref{eq-qubitone-unitary-processout-measprob-first})
yields
\begin{eqnarray}
E
\{
\Yketbasisprobrandindexzero
\}
&
=
&
\cos
(
\Yopmixparamthree
)
E
\{
\Yketbasiscoefrandindexzeroamongtwomod
^2
\}
+
\frac{
1
-
\cos
(
\Yopmixparamthree
)
}
{2}
\label{eq-qubitone-unitary-processout-measprob-first-expect}
\\
&
&
-
\cos
(
\Yopmixparamfour
)
\sin
(
\Yopmixparamthree
)
E
\{
\cos
(
\Yketbasiscoefrandindexoneamongtwophase
)
\}
E
\{
\Yketbasiscoefrandindexzeroamongtwomod
\sqrt{
1
-
\Yketbasiscoefrandindexzeroamongtwomod
^2
}
\}
.
\nonumber
\end{eqnarray}
Eq.
(\ref{eq-qubitone-unitary-trace-vs-prob})
and
(\ref{eq-qubitone-unitary-processout-measprob-first-expect})
lead to the following conclusions.
First,
$
\Ytrace(\Yopdensity\Yobservableopstdone)
$
dos not depend on
\yopmixparamone,
as expected from Section
\ref{sec-appli-system}.
Besides,
$
\Ytrace(\Yopdensity\Yobservableopstdone)
$
turns out not to depend on
\yopmixparamtwo,
due to the considered type of measurements
(and this is also true for
\yketbasisprobrandindexzero\
itself,
not only for its expectation,
as shown by
(\ref{eq-qubitone-unitary-processout-measprob-first})).
Therefore, the approach considered here cannot estimate
\yopmixparamone\
and
\yopmixparamtwo.
Finally, by deriving an estimate of the mean value
$
\Ytrace(\Yopdensity\Yobservableopstdone)
$
from measurements,
(\ref{eq-qubitone-unitary-trace-vs-prob})
and
(\ref{eq-qubitone-unitary-processout-measprob-first-expect})
only provide a
\emph{single} equation with \emph{two} unknowns,
namely
\yopmixparamthree\
and
\yopmixparamfour\
(the required statistics of
\yketbasiscoefrandindexzeroamongtwomod\
and
\yketbasiscoefrandindexoneamongtwophase\
are known, as explained in Section
\ref{sec-appli-method-HOS}).
This
single equation
is therefore not sufficient for deriving the values of these
two unknowns, so that this approach fails to solve the considered
problem.
In contrast, we will now show that our 
approach
to RCPS 
based on 
probability statistics
succeeds in estimating
\yopmixparamthree\
and
\yopmixparamfour\
from the same type of measurement results
as in the method considered here,
by further exploiting these classical-form data.
\subsection{Approach based on the 
statistics of the
random probability
\yketbasisprobrandindexzero}
\label{sec-appli-method-HOS}
We here propose an approach that is based on the
principles introduced in Section
\ref{sec-exploit-hos}
and that therefore exploits statistical parameters
of the RV
\yketbasisprobrandindexzero.
As shown by
(\ref{eq-qubitone-unitary-processout-measprob-first}),
this RV and hence its statistical parameters only depend on
\yopmixparamthree\
and
\yopmixparamfour,
not on
\yopmixparamone\
and
\yopmixparamtwo.
Therefore,
we here only aim at estimating
\yopmixparamthree\
and
\yopmixparamfour\
(see the above comment 
about 
the possible use of other types of
measurements to estimate
\yopmixparamtwo).
To this end,
we consider
two statistical parameters of
\yketbasisprobrandindexzero,
in order to define two (independent)
equations with unknowns
\yopmixparamthree\
and
\yopmixparamfour.
Focusing on the simplest parameters,
we first again consider the first-order
moment
(\ref{eq-qubitone-unitary-processout-measprob-first-expect})
of
\yketbasisprobrandindexzero.
In addition, we here use
its
second-order
moment,
that is,
$
E
\{
(
\Yketbasisprobrandindexzero
)
^2
\}
$
\footnote{Using the variance of
\yketbasisprobrandindexzero\
instead would be equivalent, as shown by
(\ref{eq-variance}).}.
Due to
(\ref{eq-statepurerand-measprob}),
with respect to the random coefficient
\yketbasiscoefrandindexzero\
of the considered quantum state,
the statistical parameters
used here
are thus
$
E
\{
|
\Yketbasiscoefrandindexzero
|
^2
\}
$
and
$
E
\{
|
\Yketbasiscoefrandindexzero
|
^4
\}
$,
i.e.
second-order and fourth-order 
statistics of
the RV
\yketbasiscoefrandindexzero.

Considering the same conditions as in Section
\ref{sec-appli-method-mean-value},
the expression of
$
E
\{
(
\Yketbasisprobrandindexzero
)
^2
\}
$
with respect to
\yopmixparamthree\
and
\yopmixparamfour\
is derived from
(\ref{eq-qubitone-unitary-processout-measprob-first}).
Then
substituting
\yopmixparamfour\
thanks to
(\ref{eq-qubitone-unitary-processout-measprob-first-expect})
yields
\begin{equation}
E
\{
(
\Yketbasisprobrandindexzero
)
^2
\}
=
\Yketbasisprobrandindexzeropowercoeftwo
\cos
^2
(
\Yopmixparamthree
)
+
\Yketbasisprobrandindexzeropowercoefone
\cos
(
\Yopmixparamthree
)
+
\Yketbasisprobrandindexzeropowercoefzero
\label{eq-opmixparamthree-cos-eq-pol-two}
\end{equation}
with
\begin{eqnarray}
\Yketbasisprobrandindexzeropowercoeftwo
&
=
&
\frac{1}{4}
+
\frac{1}{2}
\left(
E
\{
\Yketbasiscoefrandindexzeroamongtwomod
^2
\}
-
E
\{
\Yketbasiscoefrandindexzeroamongtwomod
^4
\}
\right)
\left(
E
\{
\cos
(
2
\Yketbasiscoefrandindexoneamongtwophase
)
\}
-
3
\right)
\nonumber
\\
&
&
+
2
\Yketbasisprobrandindexzeropowercoefsubtermone
\left(
E
\{
\Yketbasiscoefrandindexzeroamongtwomod
^2
\}
-
\frac{1}{2}
\right)
+
\Yketbasisprobrandindexzeropowercoefsubtermtwo
\left(
E
\{
\Yketbasiscoefrandindexzeroamongtwomod
^2
\}
-
\frac{1}{2}
\right)
^2
\label{eq-def-ketbasisprobrandindexzeropowercoeftwo}
\\
\Yketbasisprobrandindexzeropowercoefone
&
=
&
\left(
1
-
2
E
\{
\Yketbasisprobrandindexzero
\}
\right)
\left(
\Yketbasisprobrandindexzeropowercoefsubtermone
+
\Yketbasisprobrandindexzeropowercoefsubtermtwo
\left(
E
\{
\Yketbasiscoefrandindexzeroamongtwomod
^2
\}
-
\frac{1}{2}
\right)
\right)
\\
\Yketbasisprobrandindexzeropowercoefzero
&
=
&
E
\{
\Yketbasisprobrandindexzero
\}
-
\frac{1}{4}
+
\Yketbasisprobrandindexzeropowercoefsubtermtwo
\left(
E
\{
\Yketbasisprobrandindexzero
\}
-
\frac{1}{2}
\right)
^2
\nonumber
\\
&
&
+
\frac{1}{2}
\left(
1
-
E
\{
\cos
(
2
\Yketbasiscoefrandindexoneamongtwophase
)
\}
\right)
\left(
E
\{
\Yketbasiscoefrandindexzeroamongtwomod
^2
\}
-
E
\{
\Yketbasiscoefrandindexzeroamongtwomod
^4
\}
\right)
\end{eqnarray}
where
\begin{eqnarray}
\Yketbasisprobrandindexzeropowercoefsubtermone
&
=
&
\frac{1}{2}
-
\frac{
E
\{
\Yketbasiscoefrandindexzeroamongtwomod
^3
\sqrt{
1
-
\Yketbasiscoefrandindexzeroamongtwomod
^2
}
\}
}
{
E
\{
\Yketbasiscoefrandindexzeroamongtwomod
\sqrt{
1
-
\Yketbasiscoefrandindexzeroamongtwomod
^2
}
\}
}
\\
\Yketbasisprobrandindexzeropowercoefsubtermtwo
&
=
&
\frac{
E
\{
\cos
(
2
\Yketbasiscoefrandindexoneamongtwophase
)
\}
\left(
E
\{
\Yketbasiscoefrandindexzeroamongtwomod
^2
\}
-
E
\{
\Yketbasiscoefrandindexzeroamongtwomod
^4
\}
\right)
}
{
\left[
E
\{
\cos
(
\Yketbasiscoefrandindexoneamongtwophase
)
\}
E
\{
\Yketbasiscoefrandindexzeroamongtwomod
\sqrt{
1
-
\Yketbasiscoefrandindexzeroamongtwomod
^2
}
\}
\right]
^2
}
.
\label{eq-def-ketbasisprobrandindexzeropowercoefsubtermtwo}
\end{eqnarray}

To estimate \yopmixparamthree\
from
(\ref{eq-opmixparamthree-cos-eq-pol-two}),
the required statistical parameters of
\yketbasiscoefrandindexzeroamongtwomod\
and
\yketbasiscoefrandindexoneamongtwophase\
should be known.
This yields two
estimation methods.
In the most conventional,
i.e. non-blind, method,
one performs measurements 
for (copies of)
realizations of
the input state 
\yketrandprocessin\
of the considered process
and then derives sample statistics for
the required statistical parameters.
Instead, we hereafter focus on a blind,
hence more challenging, method,
i.e. without performing any measurements
at the input of the considered process but
only using some statistical properties
imposed 
on 
that input
\footnote{Classical Blind Source Separation (BSS) methods are sometimes
stated to be ``semi-blind'', rather than ``blind'',
because they require some 
prior knowledge about the source signals to be separated, e.g. these
signals may be requested to be statistically independent.
That term ``semi-blind'' is especially used for methods that 
are more constraining concerning that prior knowledge,
e.g. 
methods
that constrain
some source moments to be known or to belong to
known intervals
in addition to requesting source independence.
From that point of view, the basic version of the
quantum estimation method proposed hereafter might
be stated to be ``semi-blind'' because, in addition to requesting
\yketbasiscoefrandindexzeroamongtwomod\
and
\yketbasiscoefrandindexoneamongtwophase\
to be statistically independent, it 
uses
additional constraints on
the marginal statistics of
\yketbasiscoefrandindexzeroamongtwomod\
and
\yketbasiscoefrandindexoneamongtwophase,
as detailed in
Section
\ref{sec-appli-method-mean-value}
(in fact, the proposed quantum estimation
method does not require one to know all the statistical
distributions of
\yketbasiscoefrandindexzeroamongtwomod\
and
\yketbasiscoefrandindexoneamongtwophase\
but only the resulting parameters defined in
(\ref{eq-ketbasiscoefrandindexzeroamongtwomod-meanpower-expression})-%
(\ref{eq-mean-cos-two-ketbasiscoefrandindexoneamongtwophase-expression})).
Anyway, it remains that 
this
proposed quantum estimation
method does not require the \textit{individual values} of the
input to be known, which is the main feature of blind and associated
methods.}.
More precisely, since we here again use the statistical
distributions of
\yketbasiscoefrandindexzeroamongtwomod\
and
\yketbasiscoefrandindexoneamongtwophase\
defined in
Section
\ref{sec-appli-method-mean-value},
the statistical parameters of
\yketbasiscoefrandindexzeroamongtwomod\
and
\yketbasiscoefrandindexoneamongtwophase\
used 
in
(\ref{eq-def-ketbasisprobrandindexzeropowercoeftwo})-%
(\ref{eq-def-ketbasisprobrandindexzeropowercoefsubtermtwo})
may be shown to read
\begin{eqnarray}
E
\{
\Yketbasiscoefrandindexzeroamongtwomod
^2
\}
&
=
&
\frac{1}{3}
(
\Yketbasiscoefrandindexzeroamongtwomoddistribboundlow
^2
+
\Yketbasiscoefrandindexzeroamongtwomoddistribboundlow
\Yketbasiscoefrandindexzeroamongtwomoddistribboundhigh
+
\Yketbasiscoefrandindexzeroamongtwomoddistribboundhigh
^2
)
\label{eq-ketbasiscoefrandindexzeroamongtwomod-meanpower-expression}
\\
E
\{
\Yketbasiscoefrandindexzeroamongtwomod
^4
\}
&
=
&
\frac{1}{5}
(
\Yketbasiscoefrandindexzeroamongtwomoddistribboundlow
^4
+
\Yketbasiscoefrandindexzeroamongtwomoddistribboundlow
^3
\Yketbasiscoefrandindexzeroamongtwomoddistribboundhigh
+
\Yketbasiscoefrandindexzeroamongtwomoddistribboundlow
^2
\Yketbasiscoefrandindexzeroamongtwomoddistribboundhigh
^2
+
\Yketbasiscoefrandindexzeroamongtwomoddistribboundlow
\Yketbasiscoefrandindexzeroamongtwomoddistribboundhigh
^3
+
\Yketbasiscoefrandindexzeroamongtwomoddistribboundhigh
^4
)
\\
E
\{
\Yketbasiscoefrandindexzeroamongtwomod
\sqrt{
1
-
\Yketbasiscoefrandindexzeroamongtwomod
^2
}
\}
&
=
&
\frac{- 1}{
3
(
\Yketbasiscoefrandindexzeroamongtwomoddistribboundhigh
-
\Yketbasiscoefrandindexzeroamongtwomoddistribboundlow
)}
\left(
\left(
1
-
\Yketbasiscoefrandindexzeroamongtwomoddistribboundhigh
^2
\right)
^{3/2}
-
\left(
1
-
\Yketbasiscoefrandindexzeroamongtwomoddistribboundlow
^2
\right)
^{3/2}
\right)
\nonumber
\\
&
&
\\
E
\{
\Yketbasiscoefrandindexzeroamongtwomod
^3
\sqrt{
1
-
\Yketbasiscoefrandindexzeroamongtwomod
^2
}
\}
&
=
&
\frac{1}{
\Yketbasiscoefrandindexzeroamongtwomoddistribboundhigh
-
\Yketbasiscoefrandindexzeroamongtwomoddistribboundlow
}
\left[
-
\frac{1}{3}
\left(
\left(
1
-
\Yketbasiscoefrandindexzeroamongtwomoddistribboundhigh
^2
\right)
^{3/2}
-
\left(
1
-
\Yketbasiscoefrandindexzeroamongtwomoddistribboundlow
^2
\right)
^{3/2}
\right)
\right.
\nonumber
\\
&
&
\left.
+
\frac{1}{5}
\left(
\left(
1
-
\Yketbasiscoefrandindexzeroamongtwomoddistribboundhigh
^2
\right)
^{5/2}
-
\left(
1
-
\Yketbasiscoefrandindexzeroamongtwomoddistribboundlow
^2
\right)
^{5/2}
\right)
\right]
\\
E
\{
\cos
(
\Yketbasiscoefrandindexoneamongtwophase
)
\}
&
=
&
\frac{
\sin
(
\Yketbasiscoefrandindexoneamongtwophasedistribbound
)
}
{
\Yketbasiscoefrandindexoneamongtwophasedistribbound
}
\\
E
\{
\cos
(
2
\Yketbasiscoefrandindexoneamongtwophase
)
\}
&
=
&
\frac{
\sin
(
2
\Yketbasiscoefrandindexoneamongtwophasedistribbound
)
}
{
2
\Yketbasiscoefrandindexoneamongtwophasedistribbound
}
.
\label{eq-mean-cos-two-ketbasiscoefrandindexoneamongtwophase-expression}
\end{eqnarray}

Therefore, when 
\yketbasiscoefrandindexzeroamongtwomoddistribboundlow,
\yketbasiscoefrandindexzeroamongtwomoddistribboundhigh\
and
\yketbasiscoefrandindexoneamongtwophasedistribbound\
are fixed to known values
and
estimates of
$
E
\{
\Yketbasisprobrandindexzero
\}
$
and
$
E
\{
(
\Yketbasisprobrandindexzero
)
^2
\}
$
are derived from measurements,
(\ref{eq-opmixparamthree-cos-eq-pol-two})
yields a second-order polynomial equation with respect to
$
\cos
(
\Yopmixparamthree
)
$.

The corresponding solutions for
$
\Yopmixparamthree
\in
[
-
\pi
,
\pi
]
$
read
\begin{eqnarray}
\Yopmixparamthree
&
=
&
\Yopmixparamthreesolindetermsigntwo
\
\mbox{arccos}
\left(
\frac{
-
\Yketbasisprobrandindexzeropowercoefone
+
\Yopmixparamthreesolindetermsignone
\sqrt{
\Yketbasisprobrandindexzeropowercoefone
^2
-
4
\Yketbasisprobrandindexzeropowercoeftwo
\left(
\Yketbasisprobrandindexzeropowercoefzero
-
E
\{
(
\Yketbasisprobrandindexzero
)
^2
\}
\right)
}
}
{
2
\Yketbasisprobrandindexzeropowercoeftwo
}
\right)
\nonumber
\\
\label{eq-opmixparamthree-sol}
\end{eqnarray}
with
$
\Yopmixparamthreesolindetermsignone
=
\pm
1
$
and
$
\Yopmixparamthreesolindetermsigntwo
=
\pm
1
$.
The value of
$
\Yopmixparamfour
\in
[
- \pi
,
\pi
]
$
is then derived from
(\ref{eq-qubitone-unitary-processout-measprob-first-expect}),
which yields
\begin{eqnarray}
\Yopmixparamfour
&
=
&
\Yopmixparamfoursolindetermsign
\mbox{arccos}
\left(
\frac{
- E
\{
\Yketbasisprobrandindexzero
\}
+
\cos
(
\Yopmixparamthree
)
E
\{
\Yketbasiscoefrandindexzeroamongtwomod
^2
\}
+
\frac{
1
-
\cos
(
\Yopmixparamthree
)
}
{2}
}
{
\sin
(
\Yopmixparamthree
)
E
\{
\cos
(
\Yketbasiscoefrandindexoneamongtwophase
)
\}
E
\{
\Yketbasiscoefrandindexzeroamongtwomod
\sqrt{
1
-
\Yketbasiscoefrandindexzeroamongtwomod
^2
}
\}
}
\right)
\nonumber
\\
\label{eq-opmixparamfour-sol}
\end{eqnarray}
with
$
\Yopmixparamfoursolindetermsign
=
\pm 1
$.

First disregarding the choice of
\yopmixparamthreesolindetermsignone,
\yopmixparamthreesolindetermsigntwo\
and
\yopmixparamfoursolindetermsign,
the main result thus obtained is that
(\ref{eq-opmixparamthree-sol})
and
(\ref{eq-opmixparamfour-sol})
show that our approach succeeds in
estimating
\yopmixparamthree\
and
\yopmixparamfour.
We again stress that this is achieved by
using 
$
E
\{
(
\Yketbasisprobrandindexzero
)
^2
\}
$,
i.e.
the statistics of
\yketbasisprobrandindexzero\
beyond the first order and hence the statistics
of 
\yketbasiscoefrandindexzero\
beyond the second order.
In constrast, by only using 
second-order
statistics of
\yketbasiscoefrandindexzero,
the approach based on the density operator
and the associated mean of measurements
$\Ytrace(\Yopdensity\Yobservableopstdone)$
fails to estimate
\yopmixparamthree\
and
\yopmixparamfour,
as shown 
in Section
\ref{sec-appli-method-mean-value}.

In the basic version of the method proposed here,
estimates of
\yopmixparamthree\
and
\yopmixparamfour\
are obtained up to some so-called indeterminacies,
corresponding to the fact that this method does not
define whether each of the parameters
\yopmixparamthreesolindetermsignone,
\yopmixparamthreesolindetermsigntwo\
and
\yopmixparamfoursolindetermsign\
should be set to 1 or
$
- 1
$.
Various types of indeterminacies also exist in
classical BSS and blind system/mixture identification,
due to the limited information available in
\textit{blind} methods.
Part of these indeterminacies
can 
e.g.
be avoided by requesting some
additional prior knowledge, that would here e.g.
correspond to knowing to which intervals the
unknown values of
\yopmixparamthree\
and
\yopmixparamfour\
belong.
Indeterminacies also appeared in 
the basic version of our previous
BQPT 
\cite{amoi6-118}
and 
BHPE
methods
\cite{amoi-ieee-tqe-2021}.
We succeeded in removing them in refined versions
of our methods, where we used additional
occurrences of the same type of
measurements, but
with different statistics for the input quantum
states.
One 
\ytextmodifartitionehundredsixtyfivevonestepone{might}
also 
investigate the use of such measurements in order to
remove the indeterminacies on
\yopmixparamthree\
and
\yopmixparamfour\ here,
if one would like to solve this problem
completely, i.e. beyond the above illustration
of the general capabilities
of higher-order statistics of
\yketbasiscoefrandindexzero.
\subsection{Test results}
\label{sec-test-results}
To validate the blind method of Section
\ref{sec-appli-method-HOS}
and to evaluate its accuracy,
we performed numerical tests
with
data derived from
a software simulation of the 
considered configuration.
Each elementary
test
consists of 
the following
stages.
We first
create
a set of
\ytwoqubitseqnb\
realizations
of the
random-coefficient pure input state
\yketrandprocessin\
defined by
the right-hand term of
(\ref{eq-statepurerandonequbit}).
Each of these 
\ytwoqubitseqnb\
realizations
is obtained
by randomly drawing 
the parameters
\yketbasiscoefrandindexzeroamongtwomod\
and
\yketbasiscoefrandindexoneamongtwophase\
and then using
(\ref{eq-statepurerandonequbit}).
We then
transfer each such realization of 
\yketrandprocessin\
through the quantum process to be identified.
This
corresponds to using
(\ref{eq-vector-coef-out-vs-in-onequbit})
with a given value of the matrix
\yopmix\
defined by
(\ref{eq-def-opmix-qubitone-unitary})
and hence with given values of the parameters
\yopmixparamone\
to
\yopmixparamfour.
This
yields 
\ytwoqubitseqnb\
realizations of 
the set of
coefficients
\yketbasiscoefrandindexstd\ of the state
\yketrandprocessout.
Besides,
we eventually
use simulated measurements 
associated with these states%
, as defined in Section
\ref{sec-appli-state-and-meas}.
For each of the
\ytwoqubitseqnb\
realizations of the 
set of
coefficients
\yketbasiscoefrandindexstd,
Eq.
(\ref{eq-statepurerand-measprob})
yields the corresponding realization of the probability
\yketbasisprobrandindexzero,
which is
used as follows.
We use \ywritereadonestatenb\
prepared copies of
the considered 
realization of the state
\yketrandprocessin\
to
simulate \ywritereadonestatenb\
random-valued
measurements,
drawn with the above value of the probability
\yketbasisprobrandindexzero.
We then derive the 
sample 
frequency,
over
these
\ywritereadonestatenb\
measurements,
of the measurement 
result
associated with
the ket
$
|
0
\rangle
$.
This sample frequency
is an
estimate
of
the considered realization of
\yketbasisprobrandindexzero.
Then
computing the average 
of these 
\ywritereadonestatenb-preparation
estimates, over all
\ytwoqubitseqnb\
realizations of the states
\yketrandprocessin\
and
hence
\yketrandprocessout,
yields
an
$(
\Ytwoqubitseqnb
\Ywritereadonestatenb
)$-preparation
estimate
of the probability expectation
$
E
\{
\Yketbasisprobrandindexzero
\}
$.
Similarly, the 
mean of the \textit{squares of}
the estimates of all
\ytwoqubitseqnb\
realizations of
\yketbasisprobrandindexzero\
yields an estimate of
$
E
\{
(
\Yketbasisprobrandindexzero
)
^2
\}
$.
Both expectation estimates
are then used by our 
quantum parameter estimation
method defined in Section
\ref{sec-appli-method-HOS},
to derive
estimates of
\yopmixparamthree\
and
\yopmixparamfour.

As an example, the parameters of the matrix
\yopmix\
of
(\ref{eq-def-opmix-qubitone-unitary})
to be identified
were set to
the same values in all tests, namely
$
\Yopmixparamone
=
\pi
/
10
$,
$
\Yopmixparamtwo
=
2
\Yopmixparamone
$,
$
\Yopmixparamthree
=
3
\Yopmixparamone
$
and
$
\Yopmixparamfour
=
4
\Yopmixparamone
$.
Besides,
the 
RV
\yketbasiscoefrandindexzeroamongtwomod\
and
\yketbasiscoefrandindexoneamongtwophase\
that define the
input state of the considered process
(see 
(\ref{eq-vector-coef-out-vs-in-onequbit}))
were uniformly drawn,
respectively over the intervals
$
[
0
,
1
]
$
and
$
[
-
\pi
/
4,
\pi
/
4
]
$.
The above
parameters 
\ytwoqubitseqnb\
and
\ywritereadonestatenb\
were varied as described further in this section.
For each considered set of conditions defined by the values of
\ytwoqubitseqnb\
and
\ywritereadonestatenb,
we performed
100 above-defined elementary
tests, with different sets of 
realizations of the state
\yketrandprocessin,
in order to assess
the statistical performance of the considered
estimation
method
over 
100 estimations of the same set
$\{
\Yopmixparamthree
,
\Yopmixparamfour
\}$
of parameter values.

\begin{figure*}
\begin{center}
\includegraphics[width=\linewidth]{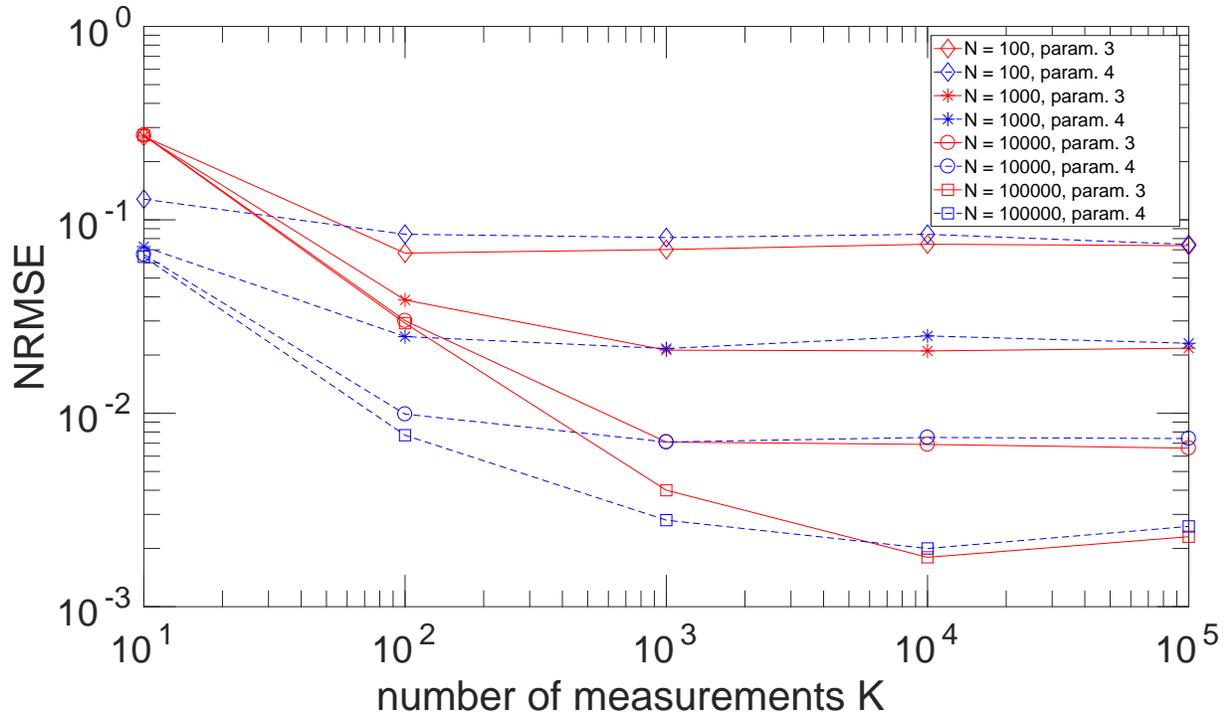}    
\caption{%
Estimation of parameters no. 3 and 4 of the matrix in
(\ref{eq-def-opmix-qubitone-unitary}),
that is,
\yopmixparamthree\
and
\yopmixparamfour:
Normalized Root Mean Square
Error (NRMSE) of
estimation
vs. number
\ywritereadonestatenb\
of 
measurements for (copies of)
each of the
\ytwoqubitseqnb\
used
states.
} 
\label{fig-id-estunitsev-bmom1mom2-plot-nrmse-param3-param4}
\end{center}
\end{figure*}

The considered performance criteria are defined as follows.
Separately for each of the parameters
\yopmixparamthree\
and
\yopmixparamfour,
we computed the Normalized Root Mean Square Error (NRMSE) 
of that
parameter over all 
100 obtained
estimates,
defined as the ratio of its
RMSE to its actual 
(positive) value.
The values of these two performance criteria are shown in
Fig. 
\ref{fig-id-estunitsev-bmom1mom2-plot-nrmse-param3-param4},
where
each plot corresponds to 
one of the parameters
\yopmixparamthree\
and
\yopmixparamfour\
and to
a fixed value of 
\ytwoqubitseqnb.
Each plot shows the variations of the 
considered performance criterion
vs. 
\ywritereadonestatenb.
We here use the 
values of
\yopmixparamthreesolindetermsignone,
\yopmixparamthreesolindetermsigntwo,
and
\yopmixparamfoursolindetermsign\
that yield the lowest NRMSE,
based on the considerations provided in Section
\ref{sec-appli-method-HOS}.

Fig.
\ref{fig-id-estunitsev-bmom1mom2-plot-nrmse-param3-param4}
first shows that the estimation error
decreases when 
\ywritereadonestatenb\
or
\ytwoqubitseqnb\
increase, as expected.
More precisely, each plot for a fixed
\ytwoqubitseqnb\
shows that
the NRMSE tends to an asymptotic value when
is
\ywritereadonestatenb\
sufficiently
increased.
This occurs because the (fixed number \ytwoqubitseqnb\ of) realizations of
\yketbasisprobrandindexzero\
are thus accurately estimated.
To further decrease that asymptotic value of
NRSME, one should then increase
the number of (estimated) values
of
\yketbasisprobrandindexzero\
over which averaging is performed, i.e.
the value of
\ytwoqubitseqnb,
as
confirmed by
Fig.
\ref{fig-id-estunitsev-bmom1mom2-plot-nrmse-param3-param4}.
This figure moreover shows that the proposed method can
achieve quite low NRMSE values, e.g. around
$
2
\times
10^{-3}
$
in the considered range of values of
\ytwoqubitseqnb\
and
\ywritereadonestatenb.
\section{Comparing discrete 
RCPS
with
mixed states}
\label{sec-compare-mixed-state}
The RV
\yketbasiscoefrandindexstd,
and hence the RV
\yketbasisprobrandindexstd\
derived from
(\ref{eq-statepurerand-measprob}),
may be continuous-valued or discrete-valued.
For instance, in the QIP problem analyzed in Section
\ref{sec-appli},
the considered
\yketbasiscoefrandindexstd\
and
\yketbasisprobrandindexstd\
are those of the output state
\yketrandprocessout\
of the process,
and the nature
(continuous/discrete)
of their statistics results from the nature of the statistics
used for 
preparing the 
random 
input state
\yketrandprocessin,
i.e. for drawing its 
parameters
\yketbasiscoefrandindexzeroamongtwomod\
and
\yketbasiscoefrandindexoneamongtwophase,
as shown by
(\ref{eq-qubitone-unitary-processout-measprob-first}).

Considering probabilistic phenomena in general,
so-called discrete
(i.e.
discrete-valued) RV are 
especially obtained
if the considered probability space
\yqipspace\
contains a finite number
\yqipoutnb\
of outcomes
\yqipout.
This then allows us to
obtain
a discrete 
RCPS 
especially
by considering a situation with
\yqipoutnb\
``possible cases'',
i.e.
\yqipoutnb\
outcomes
\yqipout,
each with a probability
of occurrence
\yqipoutprob.
Selecting
such an outcome
\yqipout\
completely defines the corresponding values of the 
coefficients 
\yketbasiscoefrandindexstdval\
of the
asssociated
pure state
\yketrandval ,
i.e. 
the value of
\yqipout\
defines that deterministic-coefficient state
\yketrandval,
that thus has a probability
\yqipoutprob.
The RV
\yketbasiscoefrandindexstd,
and hence the RV
\yketbasisprobrandindexstd\
derived from
(\ref{eq-statepurerand-measprob}),
are thus discrete.

At first sight, the above
discrete 
set of 
deterministic-coefficient pure states
\yketrandval\
and associated probabilities \yqipoutprob\
are reminiscent of how von Neumann introduces
mixed states in Chapter IV of
\cite{bookvonneumannfoundationsqm},
before he moves to their description in terms of a density
operator
(pp. 295-296).
However, the complete definition of how RCPS and mixed states are
handled moreover contains the following major difference,
which is the reason why 
they yield different properties for QIP
tasks.
When addressing
mixed states, von Neumann
considers all 
deterministic-coefficient pure states
\yketrandval\
as a whole
and he 
only
computes 
averages
of physical quantities
over all these states
\yketrandval\
(therefore involving the probabilities
\yqipoutprob),
which corresponds to only considering the quantity
$\Ytrace(\Yopdensity\Yobservableopstdone)$.
The corresponding practical procedure is based on
observable measurements, using what we here call
``unsegmented data'',
i.e. computing a single observable average over all
available data.
In contrast, as explained 
in Section
\ref{sec-prepar-measur},
our multiple-preparation
practical approach is based on 
segmented data.
This means that we require the data to be created so that,
\textit{separately for each 
outcome
\yqipout},
one accesses all measurement results
for the single state
\yketrandval.
For a given
\yqipout,
this then makes possible to estimate all probabilities
\yketbasisprobrandindexstdval\
with
$0\leq\Yindexstdforketbasis\leq\Yketspacedim-1$.
Then considering
the complete set of data differently,
separately
for any
index
\yindexstdforketbasis\
with
$0\leq\Yindexstdforketbasis\leq\Yketspacedim-1$,
we thus get
the set of
(estimated)
values
\yketbasisprobrandindexstdval\
for all
outcomes
\yqipout.
For any
\yindexstdforketbasis,
this defines 
\ytextmodifartitionehundredsixtyfivevonestepone{the whole}
statistical
distribution of
the RV
\yketbasisprobrandindexstd.
This distribution may then be exploited, thus providing
QIP capabilities that cannot be achieved when
only considering
$\Ytrace(\Yopdensity\Yobservableopstdone)$
for a mixed state.

To summarize,
richer information 
and hence better
QIP capabilities
are obtained with our random-probability-based
RCPS framework
than with mixed states
and
$\Ytrace(\Yopdensity\Yobservableopstdone)$,
but
at the expense of adding a constraint,
that is, using the above-defined segmented data of our
multiple-preparation approach:
this remains compatible with
the results that
von Neumann
obtained in a \textit{different} configuration
than ours
(unsegmented data)
and with the idea 
that 
``there is no such thing as a free lunch'',
which is reasonable.

In other words, von Neumann
defines mixed states by explicitly assuming:
``if we do not even know what state is actually present --
for example, when several states
$[...]$
with the respective probabilities
$[...]$
constitute the description'',
where the ``several states'' and
``respective probabilities''
he mentions
are
\yketrandval\
and
\yqipoutprob\
with our notations.
In practice, these mixed states are handled
by repeatedly drawing a
pure state at random,
measuring a given quantity,
and finally averaging all measurement results,
as explained above.
In our multiple-preparation 
approach to RCPS, each pure state
\yketrandval\
is also randomly drawn but, once it has been selected,
many copies of it are created 
\ytextmodifartitionehundredsixtyfivevonestepone{(as discussed in Section
\ref{sec-prepar-measur})}
and considered apart
from all the data associated with any other
deterministic-coefficient
pure state that is subsequently also
randomly drawn. 
This allows us 
to
perform averaging 
for measurements corresponding to only
the copies
of that single state
\yketrandval.
This multiple-preparation approach
thus requires 
many copies of each pure state
\yketrandval\
to accurately estimate the statistical
distributions of all RV
\yketbasisprobrandindexstd,
with
$0\leq\Yindexstdforketbasis\leq\Yketspacedim-1$.

In contrast,
we also recently developed
\textit{single-preparation} QIP methods, 
intended for BQPT, BHPE,
BQSS and related tasks,
as well as intrusion detection in quantum channels:
see details in
\cite{amoi6-104,amoi6-118,amoi-ieee-tqe-2021,amoi-qinp-2022-intrusion-detection}.
This single-preparation
approach is different from the 
multiple-preparation one but does not contradict it,
as will now be shown.
As suggested by its name, our single-preparation
approach can operate with few or even with only one
preparation of each drawn pure state
\yketrandval.
This is acceptable because, 
for any index
\yindexstdforketbasis,
we did not use this 
approach to estimate the individual probabilities
\yketbasisprobrandindexstdval\
for all
outcomes
\yqipout,
but only the expectation
$E\{\Yketbasisprobrandindexstd\}$,
i.e. the first-order moment
of the RV
\yketbasisprobrandindexstd,
using 
our procedure that we described
in
\cite{amoi6-104,amoi6-118,amoi-ieee-tqe-2021,amoi-qinp-2022-intrusion-detection}.
When we developed 
that
single-preparation
approach, we did not comment
about whether it could be extended to 
second-order and
higher-order statistics
of
\yketbasisprobrandindexstd,
but we expected
that it would be difficult,
and possibly infeasible for some of those statistical
parameters,
because some linearity properties that we used for
$E\{\Yketbasisprobrandindexstd\}$
would not hold for other parameters.
We can now extend 
that comment
by taking into
account, as follows, the considerations about
von Neumann's mixed states that we provided above.
We explained that our multiple-preparation approach
to RCPS
yields higher capabilities than the use of 
mixed states,
because it segments the measured data and it is
thus able to estimate some parameters (namely
the probabilities
\yketbasisprobrandindexstdval),
for each segment.
But when the length of each segment,
i.e. the number of copies of each state
\yketrandval,
decreases down to one,
not only the probabilities
\yketbasisprobrandindexstdval\
cannot be individually estimated,
but the concept of segment itself vanishes:
we are left we an overall set of states
\yketrandval,
with one copy of each such state,
and the only averages we can compute 
are
over this complete data set.
This
corresponds to the higher level of the
two-level procedure that we defined above,
after
(\ref{eq-statepurerand-measprob-onequbit}),
for this
multiple-preparation approach,
whereas
the lower level here disappears.
But, if only computing an overall average for the
complete set of data, we thus get back to
von Neumann's approach based on mixed states. 
Therefore, unless we will
disclose another trick for handling the
single-preparation configuration differently
for RCPS
\footnote{One may also wonder whether continuous RCPS
yield different properties than discrete ones.},
at this stage it seems that it will face the
same limitation as the approach based on mixed
states.

Two RCPS-based
approaches with complementary features 
are thus currently
available.
The first one is
the multiple-preparation approach, which has the
above-defined
advantages, that result from the use of the
second-order and higher-order statistics of the
probabilities
\yketbasisprobrandindexstd\
and the drawback of requiring
multiple and
segmented preparations.
The second one is
the single-preparation approach, which yields simpler
operation or is even required 
in some applications
(e.g. statistical intrusion detection),
as detailed 
e.g. in
\cite{amoi6-104,amoi6-118,amoi-ieee-tqe-2021,amoi-qinp-2022-intrusion-detection},
but which currently applies only to QIP problems that can be
solved by only using the expectation,
i.e. first-order statistics, of
\yketbasisprobrandindexstd.
\section{Other related works}
\label{sec-other-works-random-states}
The above-defined topics
of this investigation also compare as follows with
previous works
from the literature.
The first topic is the concept of RCPS themselves and hence its
relationships with ``random quantum states'' in a 
broad 
sense.
Of course, usual concepts
of quantum mechanics already involve randomness,
because a measurement 
performed for a deterministic-coefficient
pure state usually
yields a random result.
In the present section we do \textit{not} address that basic
type
of randomness (which corresponds to the lower level of
our multiple-preparation procedure of Section
\ref{sec-prepar-measur}),
but the 
types of randomness 
that may be defined
in addition to that basic type 
and 
to von Neuman's concepts related to mixed states that we
presented in Section
\ref{sec-compare-mixed-state}
(for our RCPS, the
additional type of randomness
corresponds to
the higher level of
our multiple-preparation procedure of Section
\ref{sec-prepar-measur}).
This yields the following three aspects.

First, 
not 
yet 
focusing on 
RCPS, some papers from the literature contain 
limited statements
about 
``random quantum states'' in a 
broad 
sense.
In particular,
\cite{amq135}
especially deals with quantum thermal states and
considers
that
``a
random state 
[...]
can be used to represent the outcome of a measurement process, 
or
to describe the statistics of an ensemble''
but does not use the concept of RCPS as defined in the present paper
(for the quantum framework,
\cite{amq135}
only mentions 
``random phases'').

Second, 
\cite{amq130}
mainly 
considers a random quantum 
pure 
state as a whole, i.e. as a vector,
without 
explicitly
providing
its mathematical expression in a given basis:
that paper is not very detailed.
It
briefly
mentions 
``the components of the
state vector, in some fixed basis''
but 
does not refer
to random variables
for these components.
Moreover, it is 
restricted to specific probability
distributions for the above quantities,
namely to the case when
``pure states are
distributed uniformly over the unit sphere''
and possibly 
in addition e.g.
``subjected to the restriction
that all the components of the state vector in the given basis be real.''
In contrast, in the present paper, we allow arbitrary probability
distributions for the ket coefficients.
This
is very important,
because it is required for being able to address
a wide range of
QIP problems,
especially blind (i.e. unsupervised)
processing problems,
where some probability distributions may be unknown.

Finally, quite a few papers,
published more recently than our first papers
\ytextmodifartitionehundredsixtyfivevonestepone{
(that include
\cite{amoi5-31}),}
have closer relationships with our work:
although they do not use the term RCPS, they use that 
concept or closely related ones,
i.e. a ket whose coefficients are random variables, or at least related
to random variables.
More precisely,
in
\cite{amq131}
the ket coefficients are defined as
``functions of complex-valued random variables
$
\boldsymbol{\xi}
$''
where
$
\boldsymbol{\xi}
$
is a vector,
whereas in
\cite{amq139}
these coefficients themselves ``are chosen at random from some given
probability distribution''.
Moreover, both
\cite{amq131}
and
\cite{amq139}
then only focus on quite specific probability distributions:
see the symmetries 
and constraints on even and odd functions
imposed in
\cite{amq131},
together with
the 
three specific
probability densities
defined in its Table I,
e.g. leading to states that are uniformly distributed over the
unit sphere;
instead
\cite{amq139}
``consider[s] the [ket coefficients]
as iid (real, complex or quaternion-real) Gaussian variables
with zero mean''
(which, by the way, cannot be an accurate model
of actual behavior:
the modulus of a ket coefficient is upper bounded by one,
so that this coefficient cannot have an unbounded Gaussian density).
In contrast, 
as stated above,
we allow arbitrary probability
distributions for the ket coefficients.

Let us then focus on the only above-mentioned
papers from the literature that are
connected with RCPS, namely
\cite{amq131,amq139}
and, to a much lower extent,
\cite{amq130}.
Those papers completely differ
from
the present one concerning its other topics,
beyond the RCPS concept.
First, the core feature analyzed 
in
this paper consists of the second-order and especially higher-order
moments of the random ket coefficients and associated random
probabilities,
including their practical estimation.
Instead,
\cite{amq131}
only mentions a very limited set of moments 
(see the three moments in Table I),
whereas
\cite{amq130,amq139}
do not mention them at all.
Second, 
\ytextmodifartitionehundredsixtyfivevonestepone{apart from}
quantum theory,
the present paper aims at
exploiting the above
moments for performing various QIP tasks, e.g. related to
QPT and quantum parameter estimation.
In contrast,
\cite{amq131}
has other goals
(quantum numerical simulation)
and
only mentions (quite a few) moments as a by-product.

Finally, we stress that some papers from the 
quantum literature mention concepts related to
higher-order moments, but 
in quite different frameworks than ours.
In particular,
in
\cite{amq136}
Mielnik
considers non-standard frameworks
as announced in his title:
``Generalized quantum mechanics''.
He especially imagines what could be done in
``hypothetical theories''
where
one would
``assume that the class of observables 
$F$ is not the set of the quadratic forms
like in orthodox theory but the set 
$F_{2n}$ of all the continuous $2n$-th order
forms''.
He
thus develops
``higher order schemes''
and
comments about
``higher order multipole moments''.
This is quite different from our approach, 
that has the following features.
We
stick to
orthodox measurements
for each deterministic pure state considered in the
lower level of our procedure,
so that each outcome probability is
equal to
(the modulus of)
a
``quadratic function of a ket coefficient''.
This relates
to Mielnik's statement:
``one might define
the orthodox quantum mechanics as a theory of such a $c$-number wave
for which only the quadratic forms are the observables''.
But, unlike Mielnik,
we
perform 
our complete set of orthodox measurements
\textit{for our
new type of states}, namely RCPS,
i.e. we organize these measurements according to the
higher level of our procedure.
Our complete approach is thus
compatible with the orthodox theory,
but yields a new feature: it
allows us
to introduce the higher-order moments associated
with the (random) ket coefficients of the considered new type of states.
Besides, Mielnik explains that
``Since the quadratic character of the observables is conditioned
by the linearity of the evolution processes the most obvious
[situation where the orthodox quantum theory would not apply]
consists in hypothetical evolution processes in which the quantum
mechanical wave function would undergo a non-linear change''.
This leads him to
``non-linear versions of quantum mechanics in which a non-linear wave
equation would play the role of the Schr\"odinger equation''.
In contrast,
our approach is fully compatible with Schr{\"o}dinger's
picture of quantum mechanics
and our
``higher-order effects''
come from the
advanced use of the statistics of random ket coefficients%
\ytextmodifartitionehundredsixtyfivevonestepone{,
allowed by the existence of RCPS themselves.}
\section{Conclusion}
\label{sec-concl}
As explained in Section
\ref{sec-density-op-of-rcps},
when considering mixed states,
von Neumann claimed that one only needs to use the
density operator
\yopdensity\
and the
mean of observable
$\Ytrace(\Yopdensity\Yobservableopstdone)$.
In the present paper, we provide a detailed 
theoretical analysis
of another type of states, that we repeatedly used in
our application-driven papers since 2007.
We call these states
``random-coefficient pure states'' or RCPS,
since their
coefficients
\yketbasiscoefrandindexstd\
in a given basis are random variables
(we 
compared 
RCPS 
with
mixed states in
Section
\ref{sec-compare-mixed-state}).
With
these RCPS too, one can associate a density
operator. 
However, restricting the use of RCPS to that
operator \yopdensity\
and moreover possibly to a mean of
observable    
$\Ytrace(\Yopdensity\Yobservableopstdone)$
would result in 
only considering the second-order statistics
of the random variables
\yketbasiscoefrandindexstd\
and therefore in
ignoring a large part of the information
available 
from
RCPS.
Instead, we proposed to exploit
the
higher-order (i.e. higher than 2) statistics of
\yketbasiscoefrandindexstd,
through the second-order and higher-order statistics of
the associated random probabilities
$
\Yketbasisprobrandindexstd
=
|
\Yketbasiscoefrandindexstd
|
^2
$.
We showed that
this allows one to
access much richer information
and 
to
solve quantum information
processing 
(QIP)
problems
that cannot be 
handled
with the 
mean value
$\Ytrace(\Yopdensity\Yobservableopstdone)$
only.
We
illustrated that phenomenon
for 
one concrete QIP
problem, related to the
well-known quantum process tomography
task.
Many other potential applications
of RCPS exist. 
Some of them were suggested
above 
and we plan to investigate such
applications in future work.

So,
having in mind Feynman's general 
statement that
``There's plenty of room at the bottom''
e.g. for computing,
we may 
summarize our main claim in
this paper as follows:
to exploit the wealth of the information available from
random-coefficient pure states,
there is plenty of
room at the higher orders
(of the statistics of 
the random coefficients
\yketbasiscoefrandindexstd\
of
these 
quantum
states).

\bibliography{biblio_yd}

\end{document}